\documentclass[aps,prl,reprint,superscriptaddress,showpacs,longbibliography,floatfix,footnoteinbib]{revtex4-1}
\usepackage[latin9]{inputenc}
\usepackage{color}
\usepackage{float}
\usepackage{bm}
\usepackage{amsmath}
\usepackage{amssymb}
\usepackage{graphicx}
\usepackage{wasysym}
\usepackage[unicode=true,
 bookmarks=false,
 breaklinks=false,pdfborder={0 0 1},backref=false,colorlinks=true]
 {hyperref}
\hypersetup{pdftitle={Title},
 plainpages=false,pdfpagelabels,linkcolor=blue,urlcolor=blue,citecolor=blue,pdfdisplaydoctitle=true,pdfduplex=DuplexFlipLongEdge}
\usepackage{minitoc}

\usepackage{minitoc}

\makeatletter

\newcommand{\lyxdot}{.}

\usepackage{units}\usepackage{wasysym}

\definecolor{orange}{rgb}{0.50, 0.20, 0.0}

\newcommand{\beginsupplement}{%
	\setcounter{page}{1}
	 \renewcommand{\thepage}{SM - \arabic{page}}%
        \setcounter{table}{0}
        \renewcommand{\thetable}{S\arabic{table}}%
        \setcounter{figure}{0}
        \renewcommand{\thefigure}{S\arabic{figure}}%
        \setcounter{section}{0}
        \renewcommand{\thesection}{S\arabic{section}}%
        \setcounter{section}{0}
        \renewcommand{\thesection}{S\arabic{section}}%
        \setcounter{subsection}{0}
        \renewcommand{\thesubsection}{S\arabic{section}.\arabic{subsection}}%
        \setcounter{equation}{0}
        \renewcommand{\theequation}{S\arabic{equation}}%

     }

\usepackage{bm}
\usepackage{braket}

\makeatother

\begin{document}
\title{Disorder driven multifractality transition in Weyl nodal loops}
\author{Miguel Gonçalves}
\affiliation{CeFEMA, Instituto Superior Técnico, Universidade de Lisboa, Av. Rovisco
Pais, 1049-001 Lisboa, Portugal}
\author{Pedro Ribeiro}
\affiliation{CeFEMA, Instituto Superior Técnico, Universidade de Lisboa, Av. Rovisco
Pais, 1049-001 Lisboa, Portugal}
\affiliation{Beijing Computational Science Research Center, Beijing 100084, China}
\author{Eduardo V. Castro}
\affiliation{Centro de Física das Universidades do Minho e Porto, Departamento
de Física e Astronomia, Faculdade de Ciências, Universidade do Porto,
4169-007 Porto, Portugal}
\affiliation{Beijing Computational Science Research Center, Beijing 100084, China}
\author{Miguel A. N. Araújo}
\affiliation{CeFEMA, Instituto Superior Técnico, Universidade de Lisboa, Av. Rovisco
Pais, 1049-001 Lisboa, Portugal}
\affiliation{Departamento de Física, Universidade de Évora, P-7000-671, Évora,
Portugal}
\affiliation{Beijing Computational Science Research Center, Beijing 100084, China}
\begin{abstract}
The effect of short-range disorder in nodal line semimetals is studied by numerically exact means. 
For arbitrary small disorder, a novel semimetallic phase is unveiled for which 
the momentum-space amplitude of the ground-state wave function 
is concentrated around the nodal line and follows a multifractal distribution. 
At a critical disorder strength, a semimetal to compressible metal transition occurs,
coinciding with a multi- to single-fractality transition. 
The universality class of this critical point is characterized by the correlation length and dynamical exponents. 
At considerably higher disorder, an Anderson metal-insulator transition takes place. Our results show that the nature of the semimetallic phase in non-clean samples is fundamentally different from a clean nodal semimetal. 
\end{abstract}
\maketitle

The robustness of certain material properties to perturbations is arguably
the most appealing property of topological matter. Topological insulators
stood out as an important class of topological materials \citep{HKrmp10,QZrmp11}
whose stability with respect to interactions and disorder is by now
fairly well established \citep{ChiuRMP2016,Rachel2018}. Gapless systems
can, however, also support non-trivial momentum-space topology and
are expected to be less robust to such effects. Among these, are the
Weyl nodal loop (WNL) semimetals, for which the valence and conduction
bands linearly touch along one-dimensional (1D) loops in the three-dimensional
(3D) momentum space \citep{Armitage2018}. Their recent theoretical
prediction \citep{Kim2015,Weng2015,PhysRevLett.115.026403} and experimental
discovery \citep{Xie2015,Bian2016} triggered intense experimental
\citep{Schoop2016,Okamoto2016,HuJin,Jin2017,Xu2017a,Lou2018,PhysRevB.99.241102,Qiu2019,Sims2019,Nakamura2019} and theoretical interest \citep{Rhim2015,Fang2015,Huang2016,Chan2016,Lu2017,Xu2017,Du2017,balents2017nodalLoop,Oroszlany2018,Martin-Ruiz2018,Wang2018,Lau2019,Ezawa2019}.

A manifestation of WNL's topological nature is the presence of surface
(``drumhead'') edge states \citep{Burkov2011,Chen2015,Weng2015,Chan2016,Zhang2016}
on surfaces parallel to the loop plane, which are induced by chiral
symmetry. Since the Fermi surface is reduced to a 1D nodal line, the
density of states (DOS), $\rho(E)$, vanishes linearly for low energies,
i.e. $\rho(E)\propto\left|E\right|$.

The robustness of the topological semimetal state to interactions
\citep{Sur2016,Nandkishore2016,Roy2017,Shapourian2018,Araujo2018}
and disorder  \citep{Syzranov2017,Chen2019} is of major importance
to understand in which conditions it might be observed. For Dirac/Weyl
systems with isolated nodal points, the effect of static disorder
has recently been addressed by a series of thorough numerical studies
\citep{Pixley2015,Pixley2016PRB,Pixley2016,PhysRevB.98.205134}. The clean-limit incompressible
semimetallic state was shown to survive up to a finite critical strength
of a box-distributed disorder potential where a transition to a compressible
diffusive metal takes place \footnote{For Gaussian distributed disorder, rare region effects were proposed to add a finite spectral weight at zero energy, causing an avoided quantum critical point \citep{Pixley2016}. However, these effects were shown to add finite spectral weight only in the neighborhood of $E=0$, not affecting the finite-disorder semimetallic phase \citep{PhysRevB.98.205134}}.

For a WNL, the exact nature of the finite disorder state is yet unknown.
Coulomb interactions were shown to induce a quasiparticle lifetime
vanishing quadratically with the excitation energy, thus yielding
Fermi liquid behavior \citep{PhysRevB.93.035138}. Weak disorder does
not change the compressibility, to leading order \citep{PhysRevB.96.115130}.
Nevertheless, disorder, with or without interactions, was found to
be marginally relevant in the clean case \citep{PhysRevB.96.115130},
pointing to a different scenario than nodal point semimetals.  Perturbative
arguments are, however, of limited use to characterize the stable
fixed point at finite disorder strength. The latter is of key importance
to understand the properties of WNL compounds, particularly with regard
to transport, which has, up to know, been assumed diffusive \citep{Mukherjee2017}.

\begin{figure}
\centering{}\includegraphics[width=1\columnwidth]{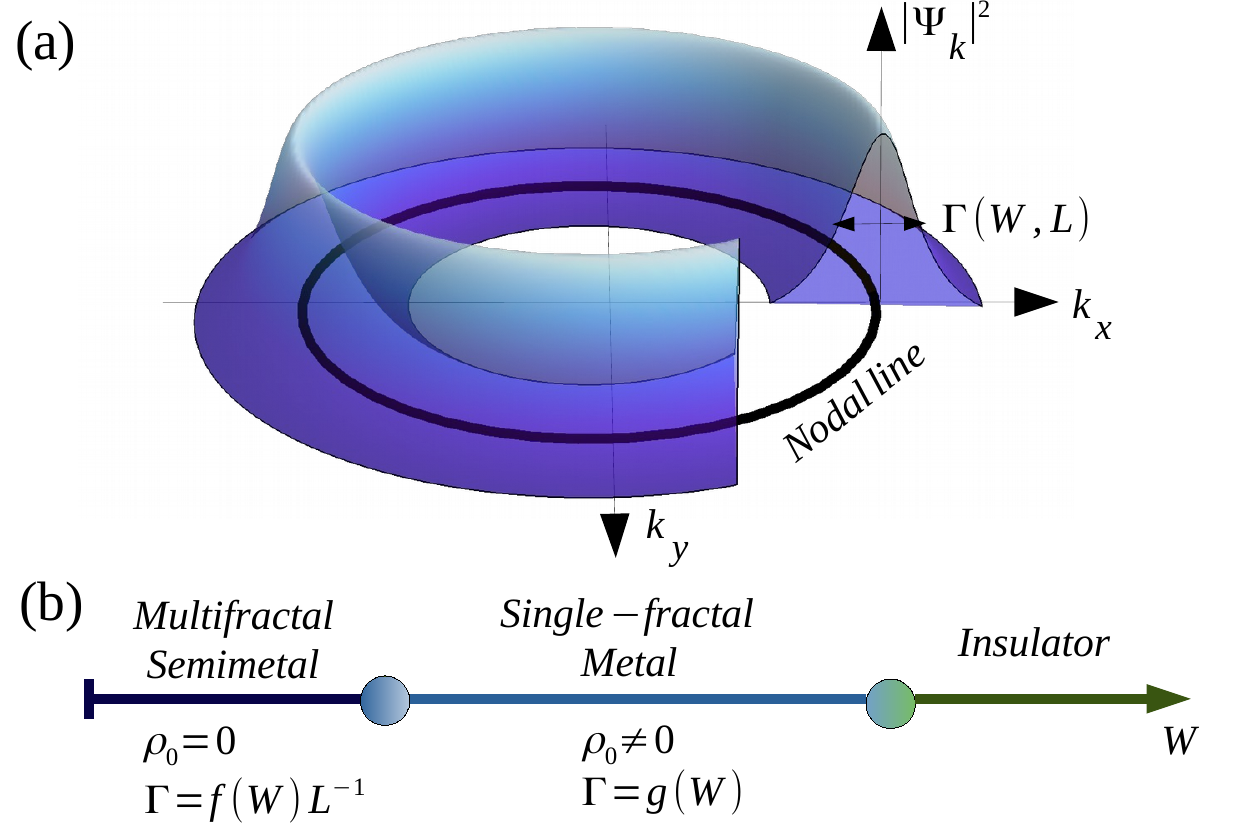}\caption{(a)~The Fermi surface of the WNL is a continuous line in the plane
$k_{z}=0$. The ground-state wave function has a width $\Gamma(W,L)$
around the loop, for fixed linear system size ($L$) and disorder
strength ($W$). (b)~Schematic phase diagram as a function of $W$.
For small $W$, the DOS at $E=0$ vanishes, $\rho_{0}=0$, and $\Gamma$
vanishes with $L^{-1}$ -- the system is in a multifractal semimetallic
phase. For $W$ larger than a critical disorder strength, $\rho_{0}\protect\neq0$
and $\Gamma$ is $L$-independent -- the system enters a single-fractal
metallic phase. For larger $W$ the system becomes an Anderson insulator.
\label{fig:General_Picture}}
\end{figure}

In this Letter, we unveil the phase diagram of a WNL in the presence
of short-range disorder using numerically exact methods. It includes
a novel multifractal (MF) semimetallic (SM) phase, corresponding to
the stable fixed point for weak disorder. Our main results are summarized
in Fig.$\,$\ref{fig:General_Picture}. We show that any small amount
of disorder mixes all the Weyl states along the nodal line depicted
in Fig.$\,$\ref{fig:General_Picture}(a), and that the width of the
wave function, $\Gamma$, vanishes as the linear system size, $L$,
increases. The resulting state is fundamentally different from the
clean one. Although the DOS still vanishes at the Fermi-level, i.e.
$\rho_{0}\equiv\rho(E=0)=0$, the momentum-space wave-function has
a multifractal structure. The MF-SM phase survives up to a critical
value of the disorder strength, where a transition to a single-fractal
(SF) metallic (M) phase takes place. In this phase the system is a
standard diffusive metal with a finite $\rho_{0}$ and $\Gamma$ loses
system size dependence. At larger disorder strength, an Anderson metal-insulator
transition occurs. The phase diagram is sketched in Fig.$\,$\ref{fig:General_Picture}(b).

\paragraph{Model and Methods.---}

We study a two-band model of a WNL on a cubic lattice with short-range
disorder,
\begin{equation}
H=\sum_{\bm{k}}c_{\bm{k}}^{\dagger}H_{\bm{k}}c_{\bm{k}}+\sum_{\bm{r}}c_{\bm{r}}^{\dagger}V_{\bm{r}}(W)c_{\bm{r}}.\label{eq:total_H}
\end{equation}
The first term describes a clean WNL, with $\bm{k}$ a 3D Bloch vector,
$H_{\bm{k}}=(t_{x}\cos(k_{x})+t_{y}\cos(k_{y})+\cos(k_{z})-m)\tau_{x}+t_{2}\sin(k_{z})\tau_{y}$,
with $\tau_{x},\tau_{y}$ Pauli matrices acting on the orbital pseudo-spin
indices $\alpha=1,2$, and $c_{\bm{k}}^{\dagger}=(\begin{array}{cc}
c_{\bm{k},1}^{\dagger} & c_{\bm{k},2}^{\dagger}\end{array})$. The second term is the disorder potential, where $\boldsymbol{r}$ is a lattice site and $V_{\bm{r}}(W)=\text{diag}(v_{\bm{r}1},v_{\bm{r}2})$,
with random variables $v_{\bm{r}\alpha}\in[-W/2,W/2]$. The results
presented hereafter are for $t_{x}=1.1$, $t_{y}=0.9$, $m=2.12$
and $t_{2}=0.8$. This choice yields a single nodal line, arising
for $k_{z}=0$. 
The hopping anisotropy breaks unwanted degeneracies and ensures the system is generic within this class.

We characterize the spectral and wave function properties by a combination
of numerical methods. To compute the DOS we use the kernel polynomial
method (KPM) with an expansion in Chebyshev polynomials to order $N_{m}$
\citep{kite_doi,Joao}, reaching system sizes up to $L=10^{3}$. To
characterize the system's lowest energy eigenstates, we use Lanczos
exact diagonalization (ED).

The eigenstates' structure is revealed by the generalized momentum-space
inverse participation ratio \citep{Pixley2018,Fu},
\begin{equation}
\mathcal{I}_{k}(q)=\Bigl(\sum_{\bm{k},\alpha}|\Psi_{\bm{k},\alpha}|^{2}\Bigr)^{-1}\sum_{\bm{k},\alpha}|\Psi_{\bm{k},\alpha}|^{2q}\propto L^{-\tau_{k}(q)},\label{eq:IPRkq}
\end{equation}

\noindent where $\Psi_{\bm{k},\alpha}$ is the eigenstate amplitude
in the $\boldsymbol{k}$ Bloch momentum state and orbital $\alpha$.
The size dependence is characterized by a $q$-dependent exponent,
$\tau_{k}$, defined in terms of the generalized dimension, $D_{k}(q)$,
as $\tau_{k}(q)=D_{k}(q)(q-1)$. In a ballistic phase, the wave function
is localized in momentum space, $\mathcal{I}_{k}(q)$ does not change
with $L$ and $D_{k}(q)=0$ for $q>0$. For a 3D-diffusive metal or
an Anderson insulator, $D_{k}(q)=3$. In these cases $D_{k}(q)$ is
constant, and the system is a single-fractal. Multifractals correspond
to cases where $D_{k}(q)$ is $q$-dependent.
This happens, for instance,  for the
real-space inverse participation ratio at a disorder driven metal-insulator
transition \citep{Janssen}.

To attenuate finite-size effects, we use twisted boundary conditions
and compute $\mathcal{I}_{k}$ averaging over random twist angles,
disorder, and the 
two lowest energy eigenstates, taking 250--1000
configurations. $\tau_{k}$ is extracted from the size dependence
of the averaged $\mathcal{I}_{k}$.

\paragraph{SM-M transition.---}

\begin{figure}
\centering{}\includegraphics[width=0.99\columnwidth]{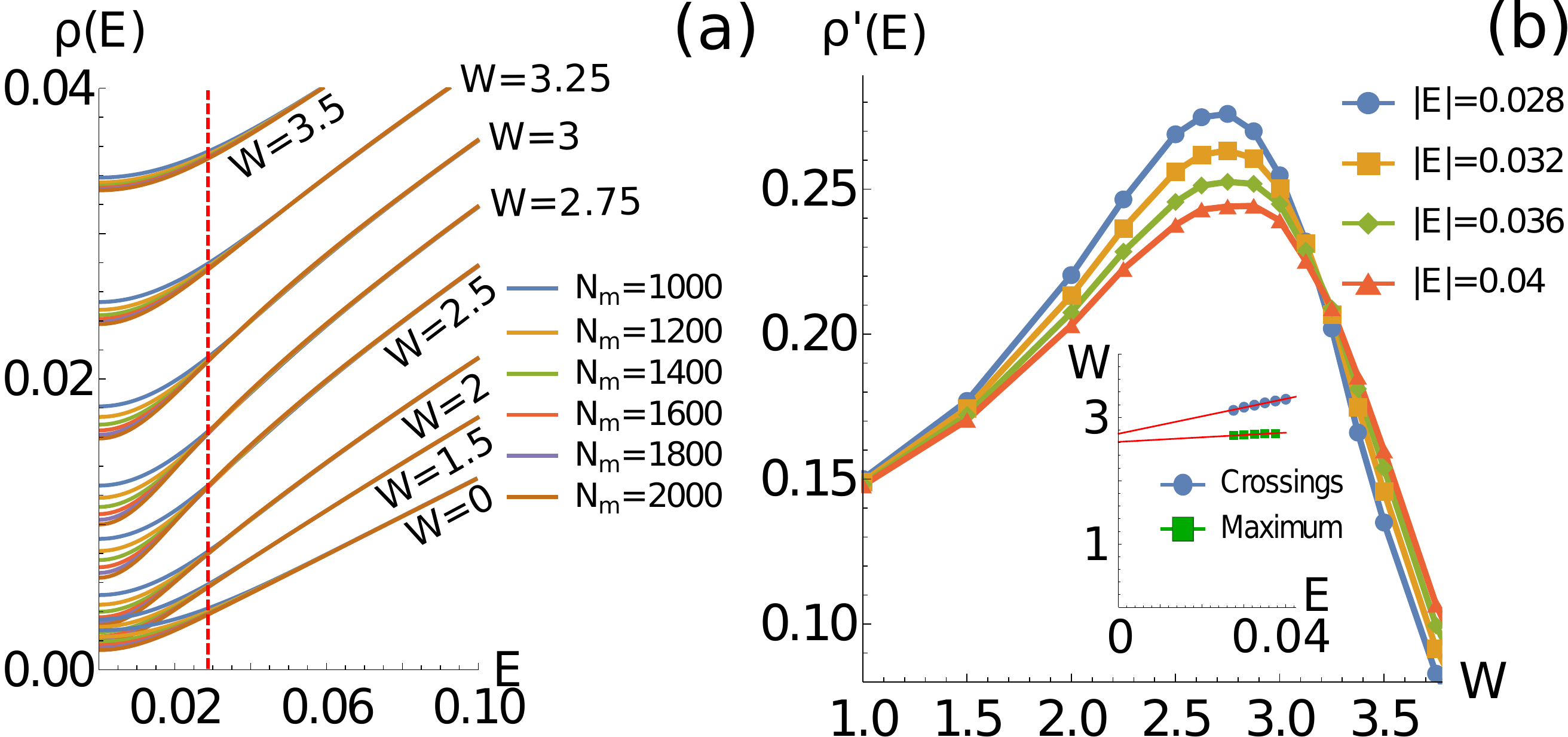}\caption{(a) DOS for different $W$ and varying $N_{m}$. For energies above
the dashed vertical line, differences between $N_{m}=1000$ and $N_{m}=2000$
are below $1\%$. (b) $\rho'(E)$, converged with $N_{m}$, as a function
of $W$, for different $E$. Inset: Extrapolation to $E\rightarrow0$
of $\rho'(E)$ crossing points ($\newmoon$) and $\rho'(E)$ maximum
($\blacksquare$). \label{fig:DOS_panel}}
\end{figure}
 The DOS for different $W$ values and varying $N_{m}$ is shown in
Fig.$\,$\ref{fig:DOS_panel}(a). Since $\rho(E)=\rho(-E)$, only
$E>0$ is plotted. For large enough $|E|$, $\rho(E)$ converges for
the highest $N_{m}$ attainable. However, within an energy window
around $E=0$, $\rho(E)$ does not converge up to the largest $N_{m}$.
This difficulty of the KPM method in resolving sharp spectral features
arises already in the clean limit and prevents a direct determination
of $\rho_{0}$ for small $W$. Nonetheless, for larger $W$ the system
is clearly metallic as $\rho_{0}$ converges to a finite value.

Quantitative predictions can be obtained from $\partial\rho/\partial E\equiv\rho'(E)$
as a function of $W$, plotted for different energies within the converged
region in Fig.$\,$\ref{fig:DOS_panel}(b). $\rho'(E)$ increases
up to a maximum value at $W=W_{{\rm max}}(E)$ and decreases abruptly
for larger $W$. Thus, there are two different regimes when $E\rightarrow0$:
for smaller (larger) $W$, $\rho'(E)$ increases (decreases) until
reaching $\rho'(0)\neq0$ ($\rho'(0)=0$). This results strongly suggest
the transition value, $W_{c}$, from a semimetal ($\rho_{0}=0$) into
a metal ($\rho_{0}\neq0$) to be  finite. In the SM phase, the growth
of $\rho'(E)$ as $E\rightarrow0$ agrees with the observed negative
concavity of $\rho(E)$ {[}see Fig.$\,$\ref{fig:DOS_panel}(a){]},
corroborating the $\rho_{0}\rightarrow0$ behavior. This provides
two ways to compute $W_{c}$: (i) Using $\lim_{E\to0}W_{{\rm max}}(E)=W_{c}$,
and extrapolating $W_{{\rm max}}(E\to0)$ from the converged region,
which yields $W_{c}=2.61\pm0.01$ {[}inset of Fig.$\,$\ref{fig:DOS_panel}(b){]};
(ii) The crossing point $W_{\text{Cross}}\left(E,\Omega\right)$,
for which $\rho'(E)=\rho'(\Omega E)$ with $\Omega>0$, obeys $\lim_{E\to0}W_{\text{Cross}}\left(E,\Omega\right)=W_{c}$.
 By computing the crossing point, $W_{\text{Cross}}\left(E,\Omega\right)$,
between $\rho'(E)$ and $\rho'(\Omega E)$ for different $E$ in the
converged region ($\Omega\simeq0.9$), we obtained a linear dependence
on $E$ and therefore we extrapolated $E\rightarrow0$ through a linear
fit, yielding $W_{c}=2.74\pm0.02$ {[}inset of Fig.$\,$\ref{fig:DOS_panel}(b){]}.

These two methods should yield the same result when $E\rightarrow0$.
However, as the lowest attainable energy is bounded by the unconverged
energy window, there is an extrapolating uncertainty in the obtained
values. We estimate the critical point by computing the least squares
between the two, yielding $W_{c}=2.64\pm0.05$, which is compatible
with the results obtained with ED \citep{sup}.

\begin{figure}
\centering{}\includegraphics[width=0.99\columnwidth]{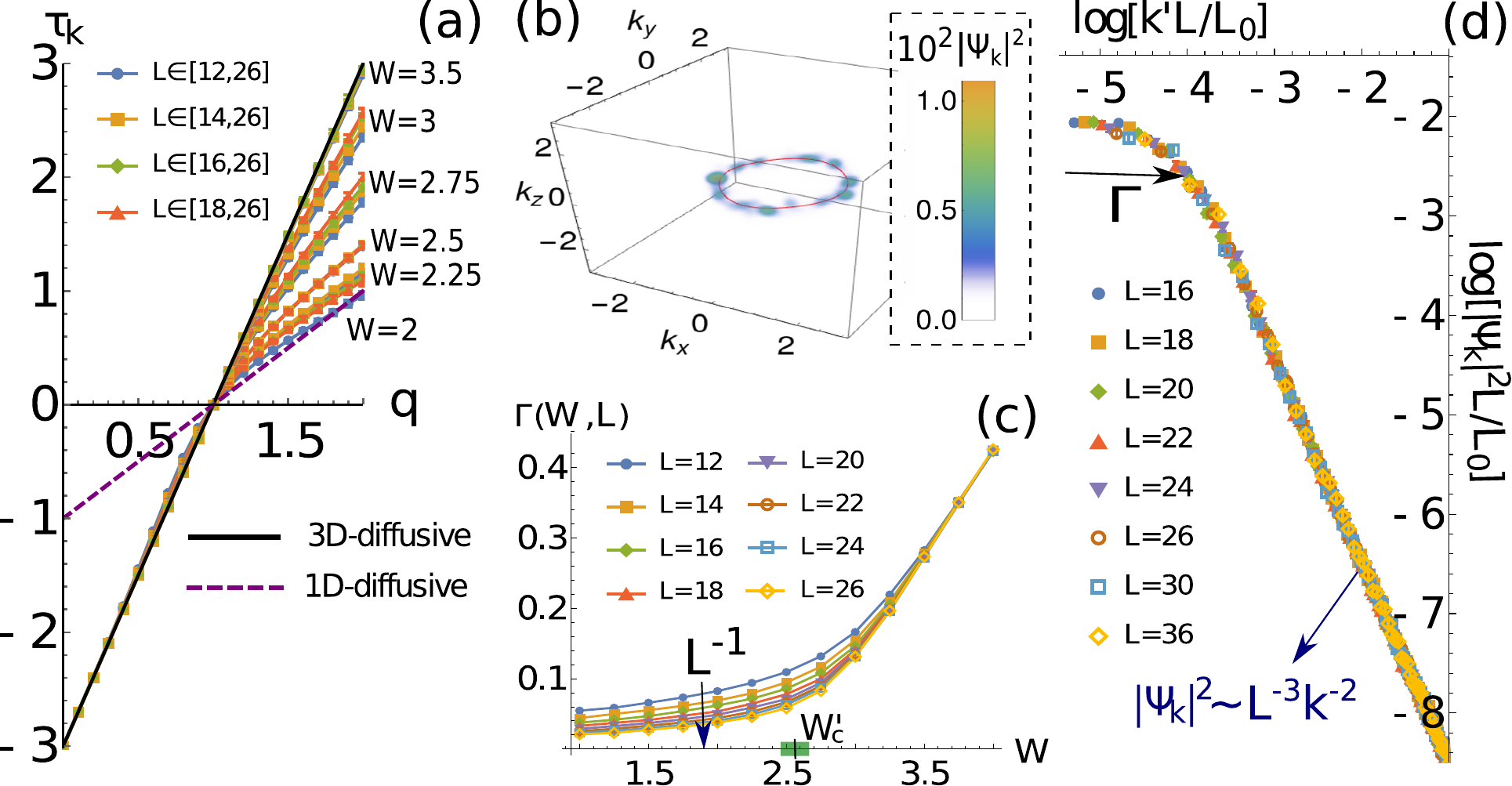}\caption{(a) Exponent $\tau_{k}(q)$ vs $q$ {[}see Eq.$\,$\ref{eq:IPRkq}{]}
for different $W$ and $L$. (b) Plot of the ground-state PD in momentum-space,
$|\Psi_{\bm{k}}|^{2}$, for a configuration with $W=3.5$. (c) $|\Psi_{\bm{k}}|^{2}$
width around the loop, $\Gamma(W,L)$, vs $W$ for varying $L$. (d)
$|\Psi_{\bm{k}}|^{2}$ as a function of $k'$ for $W=1.75$ and $k_{z}=0$,
where $k'$ is measured relative to the loop. The system is in the
MF regime and the PD curves collapse by rescaling $|\Psi_{\bm{k}}|^{2}\rightarrow|\Psi_{\bm{k}}|^{2}L$
and $k'\rightarrow k'L$. $L_{0}=16$ is the smallest used linear
system size. \label{fig:multifractal_panel}}
\end{figure}

\paragraph{MF-SF transition.---}

We now discuss the differences between the MF and SF regimes. The
computed exponent $\tau_{k}(q)$ is shown in Fig.$\,$\ref{fig:multifractal_panel}(a)
for multiple $W$ and $L$ \footnote{All the even system sizes within the $L-$intervals shown in the legend
of Fig.~\ref{fig:multifractal_panel}(a) are used to extract $\tau_{k}$.}.
 A very peculiar behavior can be observed in the MF phase: for $q<1$, $D_{k}(q)=3$, as expected for a 3D-diffusive metal;
whereas for $q>1$, $D_{k}(q)=1$, implying $\bm{k}$-space delocalization in 1D.
The origin of this phenomenon is discussed below. In the SF
case, for larger $W$, $\tau_{k}(q)$ follows the 3D-diffusive line
{[}Fig.$\,$\ref{fig:multifractal_panel}(a){]} corresponding to $D_{k}(q)=3$.
A finite size scaling analysis shows that $\tau_{k}(q)$ decreases
 (increases) with $L$ for $W<2.25$ ($W>2.75$), demonstrating the
multi (single)-fractal nature of this phase in the thermodynamic limit.
By inspection, the critical point where the MF-SF transition occurs
is thus within $W'_{c}\in]2.25,2.75[$. Below, we compute $W'_{c}$
and show it is compatible with $W_{c}$, obtained for the semimetal-metal
transition.

The origin of the MF-SF transition can be understood by inspecting
the probability distribution (PD) of the lowest energy eigenstate in
momentum space, $|\Psi_{\bm{k}}|^{2}$. As shown in Fig.$\,$\ref{fig:General_Picture}(b)
for a typical realization of disorder, the PD is concentrated along
a region of width $\Gamma$ along the nodal line. Let $\Sigma_{\text{loop}}$ be the set 
of $(\bm k,\alpha)$-points inside a torus with minor radius $\Gamma$   surrounding the WNL.
Since the loop is approximately circular, the number of points in $\Sigma_{\text{loop}}$ can be estimated as
$N\simeq2\pi\Gamma^{2}PL^{3}/(2\pi)^{3}$, where
$P$ is the loop perimeter. Since $N$ can also be estimated from $\mathcal{I}_{k}\equiv\mathcal{I}_{k}(q=2)\simeq1/N$,
we define the width of the wave function' s PD to be
\begin{equation}
\Gamma=\frac{2\pi}{\sqrt{\mathcal{I}_{k}L^{3}P}}.\label{eq:Gamma}
\end{equation}

Figure $\,$\ref{fig:multifractal_panel}(c) depicts $\Gamma$ as a
function of $W$ and $L$ \footnote{As we were only interested in the scaling of $\Gamma$ with $L$, in the plots we use $\Gamma=1/(\mathcal{I}_{k}L^{3})$.}. We found that $\Gamma\left(W,L\right)$
converges with system size in the SF phase and scales to zero with
$L^{-1}$ in the MF phase.

Within the MF phase, a scaling analysis of $|\Psi_{\bm{k}}|^{2}$
in the plane $k_{z}=0$ is shown in Fig.$\,$\ref{fig:multifractal_panel}(d).
The rescalings $|\Psi_{\bm{k}}|^{2}\rightarrow|\Psi_{\bm{k}}|^{2}L$
and $k'\rightarrow k'L$, where $k'$ is the toroidal  minor radial coordinate, make 
the numerical results for different $L$ collapse. This  shows
that, in this regime, momentum-space can be divided in two regions:
$k'<\Gamma$, where $|\Psi_{\bm{k}}|^{2}\sim L^{-1}(k')^{0}$, and
$k'>\Gamma$, where the PD decays with $k'$ as $|\Psi_{\bm{k}}|^{2}\sim L^{-3}(k')^{-2}$.
An estimation of the generalized momentum-space inverse participation
ratio yields, in the large $L$ limit, $\mathcal{I}_{k}(q)=c_{1}\sum_{\bm{k}\in\Sigma_{\text{loop}}}L^{-q}+c_{2}\sum_{\bm{k}\neq\Sigma_{\text{loop}}}L^{-3q}(k')^{-2q}=c'_{1}L^{1-q}+c'_{2}L^{3(1-q)}$,
where $c_{1},c_{2},c'_{1},c'_{2}$ are $L$-independent constants.
This explains the results of $\tau_{k}(q)$ in Fig.$\,$\ref{fig:multifractal_panel}(a)
as the scalings $L^{3(1-q)}$ and $L^{1-q}$ respectively dominate
for $q<1$ and $q>1$. In simple words, although the larger fraction
of the wave function's PD collapses in the nodal line, there is still
a finite fraction that spreads over the rest of Brillouin Zone's volume.
In the SF phase, while the asymptotic behavior $|\Psi_{\bm{k}}|^{2}\sim L^{-3}k'^{-2}$
is also observed, the scaling collapse is obtained for $|\Psi_{\bm{k}}|^{2}\rightarrow|\Psi_{\bm{k}}|^{2}L^{3}$
\citep{sup}. 

It is worth noting that, as defined in Eq.~\eqref{eq:Gamma}, $\Gamma$
can be numerically resolved only if $\Gamma\gg2\pi/L$. However, when
restricted to the plane $k_{z}=0$ , $|\Psi_{\bm{k}}|^{2}$ is still
delocalized along the loop if the area of $\Sigma_{\text{loop}}$
restricted to $k_{z}=0$, {\it  i.e.}, $\Gamma P$, is much larger than the
area of the momentum-space cell $\left(2\pi/L\right)^{2}$. This extends
the resolution computed within the $k_{z}=0$ plane to $\Gamma\gg\left(2\pi/L\right)^{2}/P$,
and allows us to study cases with $\Gamma\le2\pi/L$ in Fig.$\,$\ref{fig:multifractal_panel}(d).
For small $W\left(\lesssim1.5\right)$, we start observing $\Gamma\sim L^{-x}$,
with $1<x<2$, that we attribute to a lack of resolution for the available
system sizes \citep{sup}.

\begin{figure}
\centering{}\includegraphics[width=0.99\columnwidth]{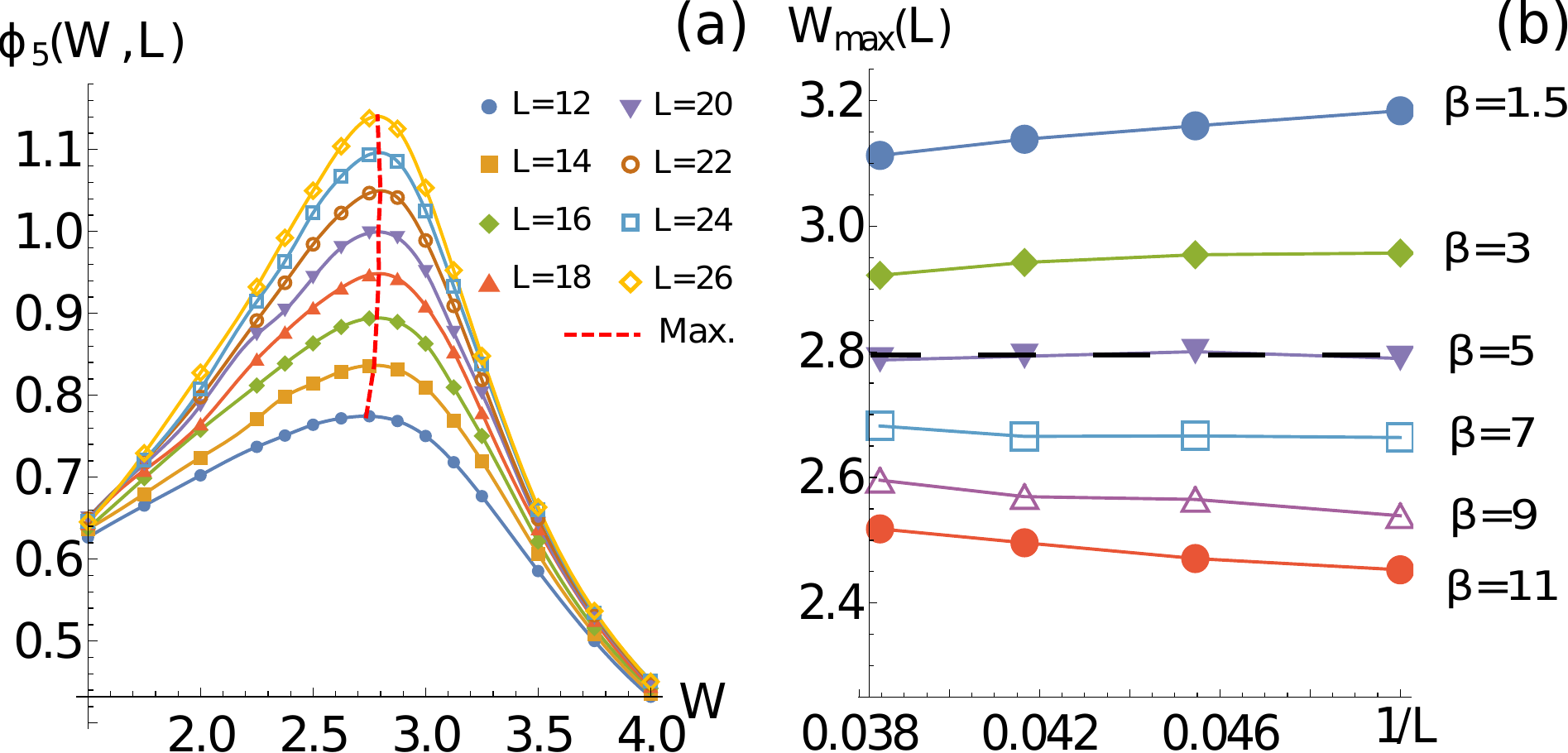}\caption{(a) The quantity $\phi_{\beta}(W,L)$ vs $W$ (see text), for $\beta=5$
and varying $L$. The red dashed line passes through the maxima of
each $L$-curve. (b) $W_{{\rm max}}(L)$ vs $L$ for different $\beta$,
with $L\in[20,26]$. The horizontal dashed, black line separates regimes
where $W_{{\rm max}}(L)$ either increases or decreases with $L$.
\label{fig4:quantity}}
\end{figure}

To estimate the critical disorder strength, $W'_{c}$,  of the MF-SF transition, we define  characteristic scales that are finite within the respective phases in the thermodynamic limit, and diverge at $W_{c}'$.
 In the MF phase, we define  $\lambda_{s}\equiv
  \Gamma LP$, 
 which diverges as $W\rightarrow W_{c}^{'-}$; 
 in the SF phase, $\lambda_{m}\equiv \Gamma^{-1}$
 diverges as $W\rightarrow W_{c}^{'+}$.
 Then, the quantity $\phi_{\beta}(W,L)=(\lambda_{s}^{-1}+\beta\lambda_{m}^{-1})^{-1}$,
with $\beta$ a positive real constant, 
 only diverges at $W=W_{c}'$. Figure.$\,$\ref{fig4:quantity}(a) shows $\phi_{\beta}(W,L)$
as a function of $L$, for $\beta=5$ and different $L$. For a fixed
$L$, $\phi_{\beta}(W,L)$ has a maximum at $W=W_{{\rm max}}(L,\beta)$.
The critical disorder strength can thus be obtained by $\lim_{L\to\infty}W_{{\rm max}}(L,\beta)=W_{c}'$,
for any $\beta>0$. However, for finite $L$ we observe a $\beta$-dependence
of $W_{{\rm max}}(L,\beta)$. As shown in Fig.$\,$\ref{fig4:quantity}(b),
there are two regimes: for $\beta<\beta_{c}\simeq5$ ($\beta>\beta_{c}$)
$W_{{\rm max}}(L,\beta)$ decreases (increases) with $L$. Thus, $W_{{\rm max}}(L,\beta_{c})$
provides an estimation of $W_{c}'$ that minimizes finite-size effects.
For $L\in[20,26]$ we find $\beta_{c}\approx5$, while for smaller
system sizes, $L\in[12,18]$, $\beta_{c}\approx3.6$. Extrapolating
$\beta_{c}$ for $L\rightarrow\infty$, we obtain $W'_{c}=2.56\pm0.10$
\citep{sup}, in good agreement with the critical value for the
SM-M transition within error bars. In the following, we take the average
value of the SM-M and MF-SF critical points and set $W_{c}=W'_{c}=2.6\pm0.1$.

\paragraph{Scaling analysis.---}

\begin{figure}
\begin{centering}
\includegraphics[width=0.99\columnwidth]{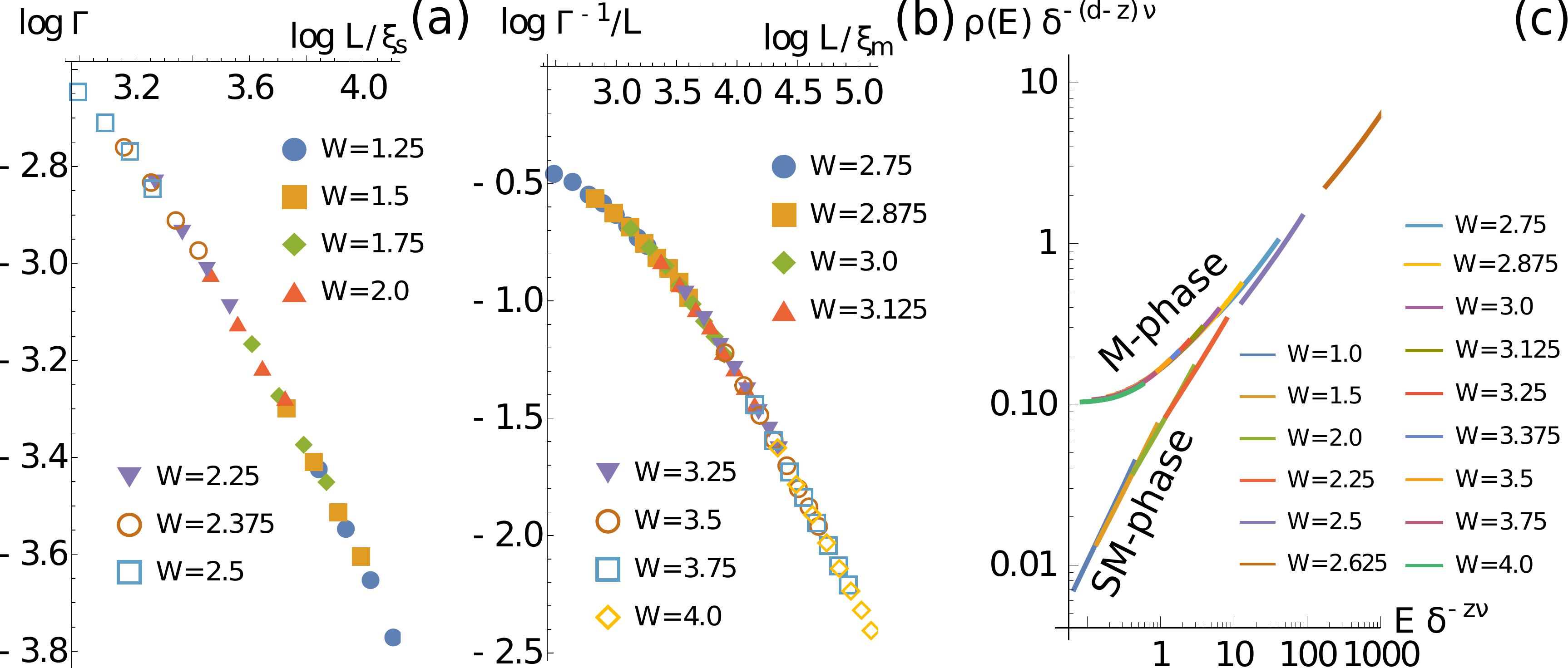}
\par\end{centering}
\caption{(a) Collapse of scaling variable $\Gamma$ by shifts of $\log\xi_{s}$,
for $L\in[20,26]$. (b) Collapse of scaling variable $\Gamma^{-1}/L$
by shifts of $\log\xi_{m}$, for $L\in[12,26]$. (c) Collapse of the
$\rho(E)$ curves according to Eq.$\,$\ref{eq:scaling_z2}, obtained
for different $W$ and $E\in[0.025,0.175]$, with parameters $\nu=1$,
$z=1.9$ and $W_{c}=2.6$. The curves collapse in two different
branches that connect at $W=W_{c}$, corresponding to the SM and M
phases.\label{fig:curve_collapsing}}
\end{figure}

We take  $\Gamma$ and $\Gamma^{-1}/L$ as finite-size scaling variables
for the SM and M phases, respectively, and write
\begin{eqnarray}
\Gamma&=&f_{s}(L/\xi_{s})\,,\\
\Gamma^{-1}/L&=&f_{m}(L/\xi_{m})\,,
\label{eq:scaling_variables}
\end{eqnarray}
\noindent where $f_{s}$ and $f_{m}$ are, respectively, scaling functions
in the SM and M phases. The thermodynamic-limit correlation lengths
$\xi_{s}$ and $\xi_{m}$, respectively in the SM and M phases, scale
as $\xi_{s},\xi_{m}\sim\delta^{-\nu}$ with $\delta=|W-W_{c}|/W_{c}$.
Collapsing the curves in Eq.$\,$\eqref{eq:scaling_variables} for
different $W$, allows the determination of $\xi_{s}$ and $\xi_{m}$
up to multiplicative constants. The data collapse is depicted in Fig.$\,$\ref{fig:curve_collapsing}(a-b).
Fitting the $\delta$ dependence as $\xi_{m}\sim\delta^{-\nu}$ yields
$\nu=1.0\pm0.2$. 
We were not able to unambiguously fit $\nu$ from $\xi_{s}$ due to the large error in its computation, arising from the resolution problems discussed before for small $W$ and finite size effects for $W$ closer to $W_c$. Nonetheless, the value of $\nu$ obtained from
$\xi_{s}$ is compatible with the scaling collapse of $\xi_{m}$ \citep{sup}.

Following Ref.$\,$\citep{Kobayashi2014}, we assume the scaling form
of the DOS near the SM-M transition to be \citep{sup}
\begin{equation}
\rho(E)\sim\delta^{\nu(d-z)}\mathcal{F}_{\gamma}(\delta^{-\nu z}|E|),\label{eq:scaling_z2}
\end{equation}
and at the transition, 
$W=W_{c},$ to vary as
\begin{equation}
\rho(E)\sim|E|^{\frac{d}{z}-1},\label{eq:scaling_z1}
\end{equation}

\noindent where the subscript $\gamma$ in Eq.$\,$\eqref{eq:scaling_z2}
distinguishes the scaling functions in the SM ($\mathcal{F}_{s}$)
and M ($\mathcal{F}_{m}$) phases. Using Eq.$\,$\eqref{eq:scaling_z1}
to fit $\rho(E)$ near $W=W_{c}$, we obtained $z=1.9\pm0.1$, where
the error is due to the uncertainty in $W_{c}$ and the variation
of the fitting energy window. This value is compatible with the results
obtained with ED \citep{sup}. Using the values of $z$ and $\nu$
determined previously, the $\rho(E)$ data collapses into two different
branches that touch at $W=W_{c}$ corresponding to the SM and M phases,
as shown in Fig.$\,$\ref{fig:curve_collapsing}(c).

As expected, the critical exponents obtained here differ from those
of the 3D metal-insulator Anderson transition (for all symmetry classes)
\citep{Wegner1976,PhysRevLett.78.4083,PhysRevLett.82.382,Asada2005},
as well as from those of a disordered Weyl semimetal ($z\approx1.5$
and $\nu\approx1$) \citep{Kobayashi2014}, confirming that this
transition belongs to a different universality class.

\paragraph{Anderson transition.---}

In the M phase, upon increasing $W$, a second phase transition takes
place at $W_{c}^{l}=11.0\pm0.2$ \citep{sup}. The critical exponent
$\nu$ is compatible with a 3D Anderson transition in the orthogonal
symmetry class, between a 3D diffusive metal and an Anderson insulator.

\paragraph{Discussion.---}

A clean WNL is unstable to an infinitesimal amount of disorder and
flows to a strong-coupling fixed point, a novel phase - here dubbed
multifractal-semimetal - where the DOS vanishes at the Fermi energy
and the momentum-space distribution of low energy states has a multifractal
structure, being concentrated on the nodal line. Upon increasing the
disorder strength, the DOS becomes finite and the eigenstate's momentum-space
distribution transitions to that of a 3D diffusive metal. Both phenomena
arise for the same critical value of disorder, $W_{c}$, up to numerical
accuracy. The ensuing multifractal semimetal to single fractal metal
phase transition belongs to a novel universality class characterized
by the critical exponents, $\nu=1.0\pm0.2$ and $z=1.9\pm0.1$,
and by the scaling functions for the DOS and correlation lengths.
Further increasing the disorder, the 3D diffusive metal transitions
to an insulating state through a phase transition of the Anderson
type.

The implications of our results to edge state physics and to the transport properties of the disordered WNL will be given elsewhere \cite{prepar}. 
It  would also be interesting to see if the rare regions effects reported for Dirac/Weyl semimetals, for gaussian-distributed disorder, do produce a finite contribution to $\rho_0$ \cite{rare1,rare2,Pixley2016} in WNL or 
otherwise leave the semimetallic phase unchanged \citep{PhysRevB.98.205134}.

We acknowledge partial support from 
Funda\c{c}\~ao para a Ci\^encia e Tecnologia (Portugal)
through Grant No. UID/CTM/04540/2013,  and UID/CTM/04540/2019. 
PR acknowledges further support through the Investigador contract IF/00347/2014.
E. V. C acknowledges partial support from FCT-Portugal through Grant No. UID/FIS/04650/2019. MG  acknowledges further support through the  Grants No. IF/00347/2014/CP1214/CT0002 and 1018P.02595.1.01 - ACTIV ID.
The hospitality of the Computational Science Research Center,
 Beijing, China, where  the final stage of this work was carried out,
 is also acknowledged.  

\bibliographystyle{apsrev4-1}
\bibliography{NodalLoops}




\clearpage
\onecolumngrid

\beginsupplement

\begin{center}
\textbf{\large Supplementary Materials for: \\
\vspace{0.1cm}
Disorder driven multifractality transition in Weyl nodal loops  }
\end{center}

\vspace{0.3cm}

In these supplemental material section we provide  additional details of our analysis and some extra numerical results.
The section is organized as follows: Sec.$\,$\ref{real_space_H} provides a real-space interpretation of the WNL's Hamiltonian;  Sec.$\,$\ref{ED_additional} gives further results on the spectral properties of the disordered WNL obtained with exact diagonalization (ED). Sec.$\,$\ref{MFSFcriticalpoint_details} presents the detailed determination of the multifractal to single-fractal critical point; Sec.$\,$\ref{critical_exps_details}  provides the details of the determination of the critical exponents $z$ and $\nu$; Sec.$\,$\ref{MItransition} is devoted to the  analysis of the metal-insulator Anderson transition; In Sec.$\,$\ref{wave_fixed_conf} we illustrate the momentum-space wave function probability for different disorder strengths.  
Finally, Sec.$\,$\ref{sec:Resolution-issues} discusses issues related to the the finite-size resolution.

\begin{center}
\vspace{0.3cm}
\par\end{center}

\vspace{0.6cm}

\twocolumngrid

\tableofcontents

\section{Real-space structure of the Hamiltonian}
\label{real_space_H}

\begin{figure}
\centering{}\includegraphics[width=0.65\columnwidth]{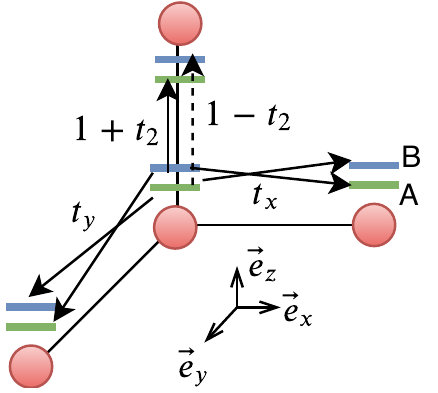}\caption{Sketch of the hopping terms of the WNL's Hamiltonian in real space
(first term in Eq.$\,$\ref{eq:total_H} of the main text). The red
circles correspond to lattice sites, and the orbitals A and B are
represented near them. The arrows represent the hopping integrals between sites.\label{fig:sketch_H}}
\end{figure}

It is useful to have a real-space representation of the Hamiltonian. 
In the clean case, it is given by the first term of Eq.$\,$(\ref{eq:total_H}), which in real space can
be written as

\begin{equation}
\begin{aligned}H= & \frac{t_{x}}{2}\sum_{i}(a_{i}^{\dagger}b_{i+l\bm{e}_{x}}+a_{i}^{\dagger}b_{i-l\bm{e}_{x}})\\
 & +\frac{t_{y}}{2}\sum_{i}(a_{i}^{\dagger}b_{i+l\bm{e}_{y}}+a_{i}^{\dagger}b_{i-l\bm{e}_{y}})\\
 & +\frac{1}{2}\sum_{i}[(1-t_{2})a_{i}^{\dagger}b_{i+l\bm{e}_{z}}+(1+t_{2})a_{i}^{\dagger}b_{i-l\bm{e}_{z}}]\\
 & -m\sum_{i}a_{i}^{\dagger}b_{i}
\end{aligned}
\end{equation}

\noindent where the sum is over real-space lattice sites, $l$ is
the lattice constant and we used $a_{i}^{\dagger}\equiv c_{i,1}^{\dagger}$
and $b_{i}^{\dagger}\equiv c_{i,2}^{\dagger}$. A sketch of the hopping
terms is provided in Fig.$\,$\ref{fig:sketch_H}.

\section{Exact diagonalization additional results (spectrum)}

\label{ED_additional}

\subsection{Scaling formulas for $\rho(E)$}

Following the arguments exposed in Ref.$\,$\citep{Kobayashi2014},
we can start by noticing that the number of states below an energy
$E$, for a system of linear size $L$ in $d$ dimensions, $\mathcal{N}(E,L)$,
should be a function of the adimensional parameters $L/\xi$ and $E/E_{0}$,
with $\xi$ and $E_{0}$ being respectively characteristic length
and energy scales:

\begin{equation}
\mathcal{N}(E,L)=f(L/\xi,E/E_{0})
\end{equation}

The dynamical exponent $z$ relates the characteristic scales $\xi$
and $E_{0}$ through $E_{0}\sim\xi^{-z}$. By using that

\begin{equation}
\rho(E)=\frac{1}{L^{d}}\frac{d\mathcal{N}(E,L)}{dE}
\end{equation}

\noindent we can write

\begin{equation}
\rho(E)=\frac{\xi^{z}}{L^{d}}f(L/\xi,E\xi^{z})=\xi^{z-d}g(L/\xi,E\xi^{z})\label{eq:rhoE_scaling_general}
\end{equation}

At the critical point, the characteristic length $\xi$ diverges and
therefore any dependence on it should be lost. This gives rise to
two scaling formulas for the DOS at the critical point. At $E=0$,
we have

\begin{equation}
\rho_{0}=\xi^{z-d}g(L/\xi,0)\sim L^{z-d},W=W_{c}\label{eq:scaling_rho0}
\end{equation}

On the other hand, in the thermodynamic limit, and using that $\rho(E)$
is an even function of $E$, we have

\begin{equation}
\rho(E)=\xi^{z-d}g(|E|\xi^{z})\sim|E|^{d/z-1},W=W_{c}\label{eq:scaling_rhoE}
\end{equation}

Finally, using that near the critical point, $\xi\sim\delta^{-\nu}$
in Eq.$\,$(\ref{eq:rhoE_scaling_general}), we get that in the thermodynamic
limit

\begin{equation}
\rho(E)\sim\delta^{\nu(d-z)}g(\delta^{-\nu z}|E|)\label{eq_Appendix:rho_E_scaling_ThermoLimit}
\end{equation}

\subsection{$W_{c}$ and critical exponent $z$}

The critical point of the semimetal-metal transition can be estimated
through exact diagonalization (ED) by studying the low energy properties
of the spectrum. We employed the Lanczos algorithm in order to compute
the lowest $N_{{\rm ev}}=24$ eigenvalues. For a given disorder strength
$W$ and system size, and for each disorder configuration, we can
compute the energies of the smallest and largest eigenvalues of the
set of $N_{{\rm ev}}$ eigenvalues, and then average over configurations
to obtain the mean energy window, $E_{w}$, of this set.

\begin{figure}[H]
\centering{}\includegraphics[width=0.9\columnwidth]{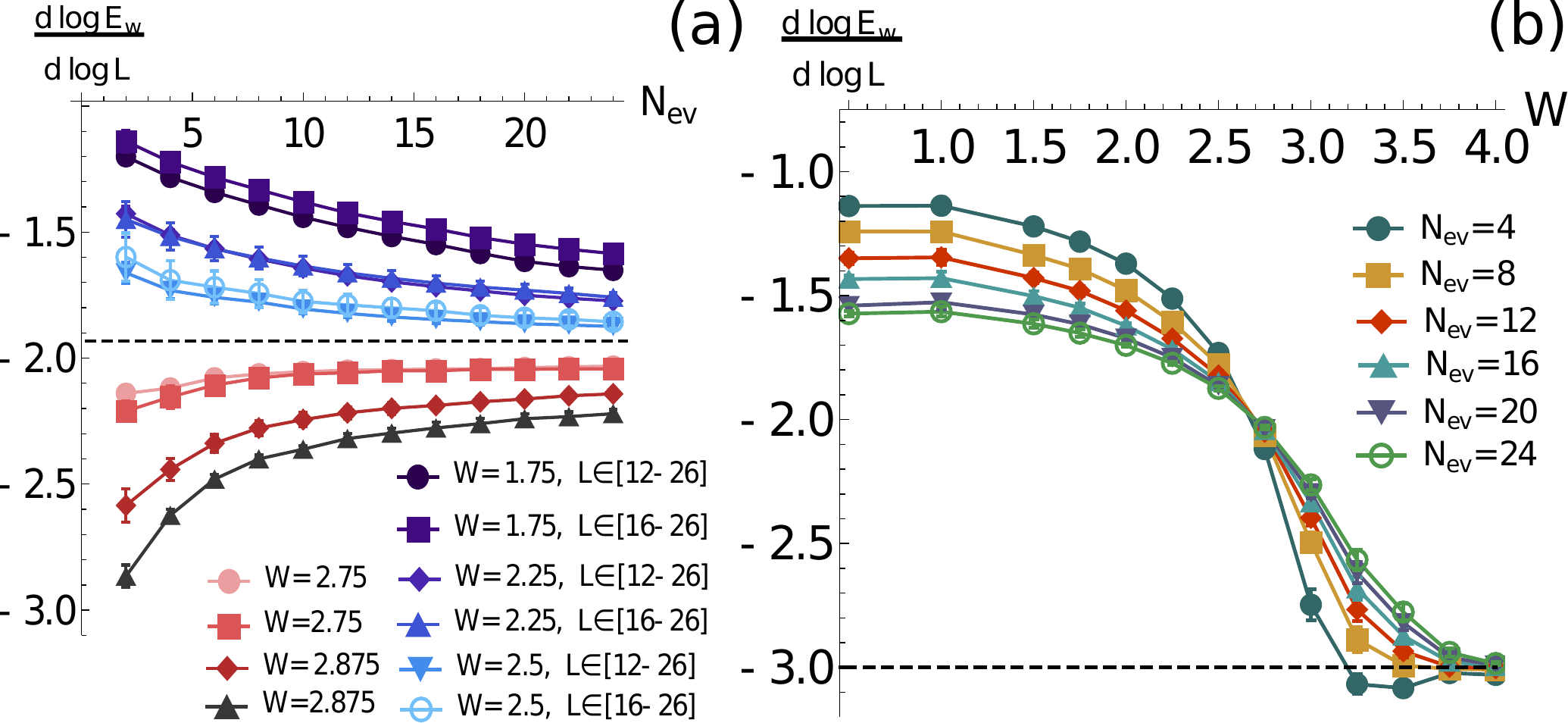}\caption{(a) $\mu\equiv d\log E_{w}/d\log L$ as a function of $N_{{\rm ev}}$
for variable $W$ and system sizes. In the legend, together with $W$,
we show the range of system sizes used in the fits to extract $\mu$
(only even $L$ was used). The dashed line represents the separation
of regimes where $\mu$ either increases or decreases with $N_{{\rm ev}}$.
(b) $\mu$ as a function of $W$ for variable $N_{{\rm ev}}$, for
$L\in[12,26]$. The crossing point occurs for $W=2.68\pm0.03$ and
$\mu=-1.98\pm0.01$. \label{fig_Appendix:Ew_method}}
\end{figure}

Important information can be extracted by studying how the energy
window $E_{w}$ scales with $L$, that is, by computing $\mu\equiv d\log E_{w}/d\log L$.
This quantity is plotted in Fig.$\,$\ref{fig_Appendix:Ew_method}.
We can see in Fig.$\,$\ref{fig_Appendix:Ew_method}(a) that there
is a qualitative change in regimes with $W$: for smaller $W$, $\mu$
decreases with $N_{{\rm ev}}$, while above some disorder strength,
the opposite is true. This translates into a crossing point as a function
of $W$, shown in Fig.$\,$\ref{fig_Appendix:Ew_method}(b), where
$\mu$ is independent of $N_{{\rm ev}}$. We will argue that this
crossing point should correspond to $W_{c}$.

For a finite system, $\rho_{0}$ can be computed
through

\begin{equation}
\rho_{0}=\langle\frac{\#N_{w}}{E_{w}L^{d}}\rangle_{N_{c}}
\end{equation}
where $\#N_{w}$ is the number of states inside the energy window
$E_{w}$ and $L^{d}$ is the system's volume. $\langle\rangle_{N_{c}}$
denotes an average over $N_{c}$ disorder configurations. By using
a fixed number of eigenvalues, $N_{{\rm ev}}$, we have  $\#N_{w}=N_{{\rm ev}}$
and that $E_{w}\sim L^{\mu}$. We must therefore have  $\rho_{0}\sim L^{-\mu-d}$
and, from Eq.(\ref{eq:scaling_rho0}),
\begin{equation}
z=-\mu\equiv-d\log E_{w}/d\log L\label{eq:z_mu}
\end{equation}

This relation could be obtained in a different way, through Eq.$\,$(\ref{eq:scaling_rhoE}).
The number of states inside an energy window $E_{w}$ should also
be

\begin{equation}
\#N_{w}\sim L^{d}\int_{0}^{E_{w}}\rho(E){\rm dE}\label{eq:Nw_rhoE}
\end{equation}

By using that $\rho(E)\sim|E|^{d/z-1}$, we have that

\begin{equation}
\#N_{w}\sim L^{d}E_{w}^{d/z}
\end{equation}

\noindent implying again Eq.$\,$(\ref{eq:z_mu}) for fixed $\#N_{w}$.

At the critical point, we must have a well defined critical exponent
$z$, independent of $N_{{\rm ev}}$. Therefore, this should be the
crossing point that we observe in Fig.$\,$\ref{fig_Appendix:Ew_method}(b),
corresponding to $z=1.98\pm0.01$ and $W_{c}=2.68\pm0.03$, compatible
with the results in the main text.

\begin{figure}[H]
\centering{}\includegraphics[width=0.99\columnwidth]{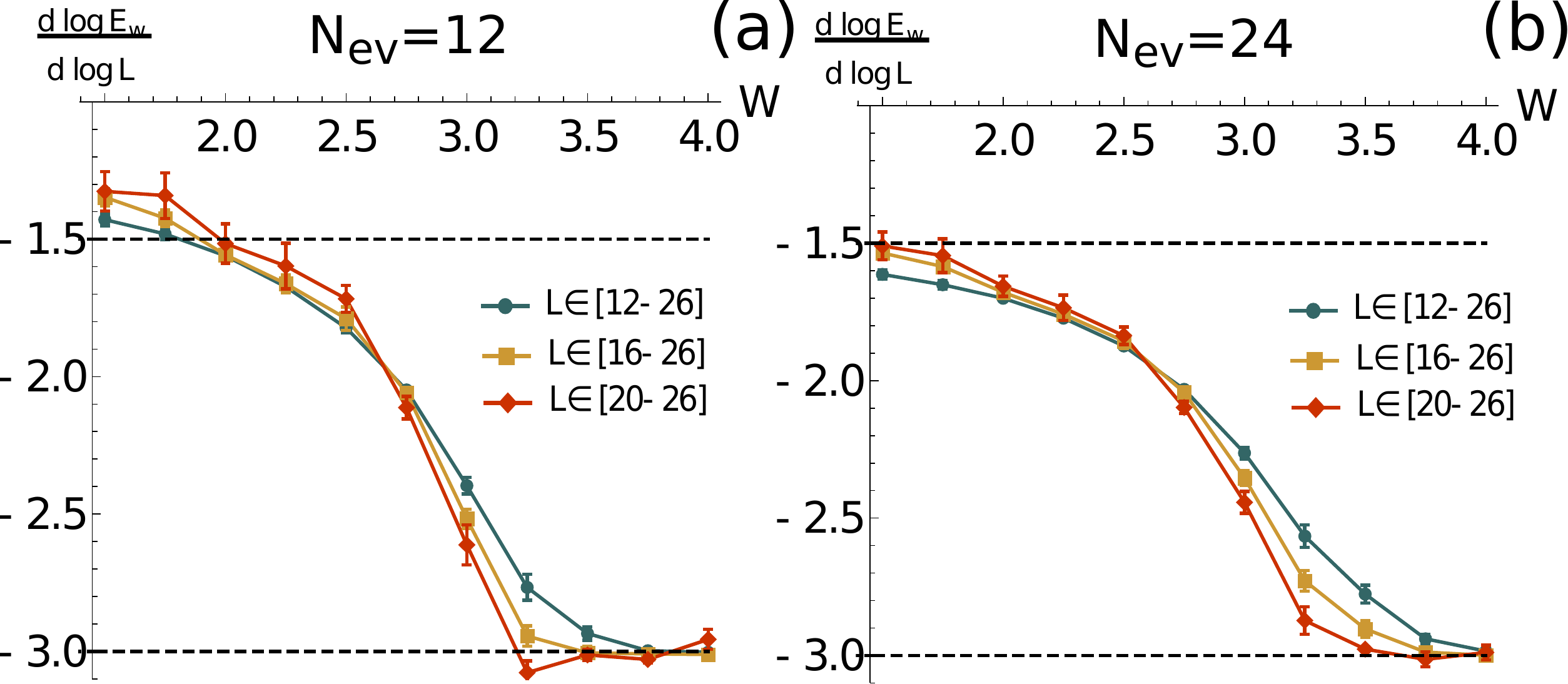}\caption{$\mu\equiv d\log E_{w}/d\log L$ as a function of $W$ for different
ranges of system sizes and fixed $N_{{\rm ev}}=12$ (a) and $N_{{\rm ev}}=24$
(b).\label{fig_Appendix:Ew_varL}}
\end{figure}

One should finally recall that the results were shown for $L\in[12,26]$.
$W_{c}$ and $z$ can, nonetheless, vary if larger systems are used.
To inspect whether there is a significant system size dependence,
we fixed $N_{{\rm ev}}$ and varied the system sizes used in the fit.
Once again crossing points were observed, matching the previously
obtained one - $W_{c}\in[2.5,2.75]$ and $z\approx1.95$. In Fig.$\,$\ref{fig_Appendix:Ew_varL},
we show results for $N_{{\rm ev}}=12$ and $N_{{\rm ev}}=24$.

\subsection{Level spacing statistics}

\label{subsec:LSS_results}

\begin{figure}[H]
\centering{}\includegraphics[width=1\columnwidth]{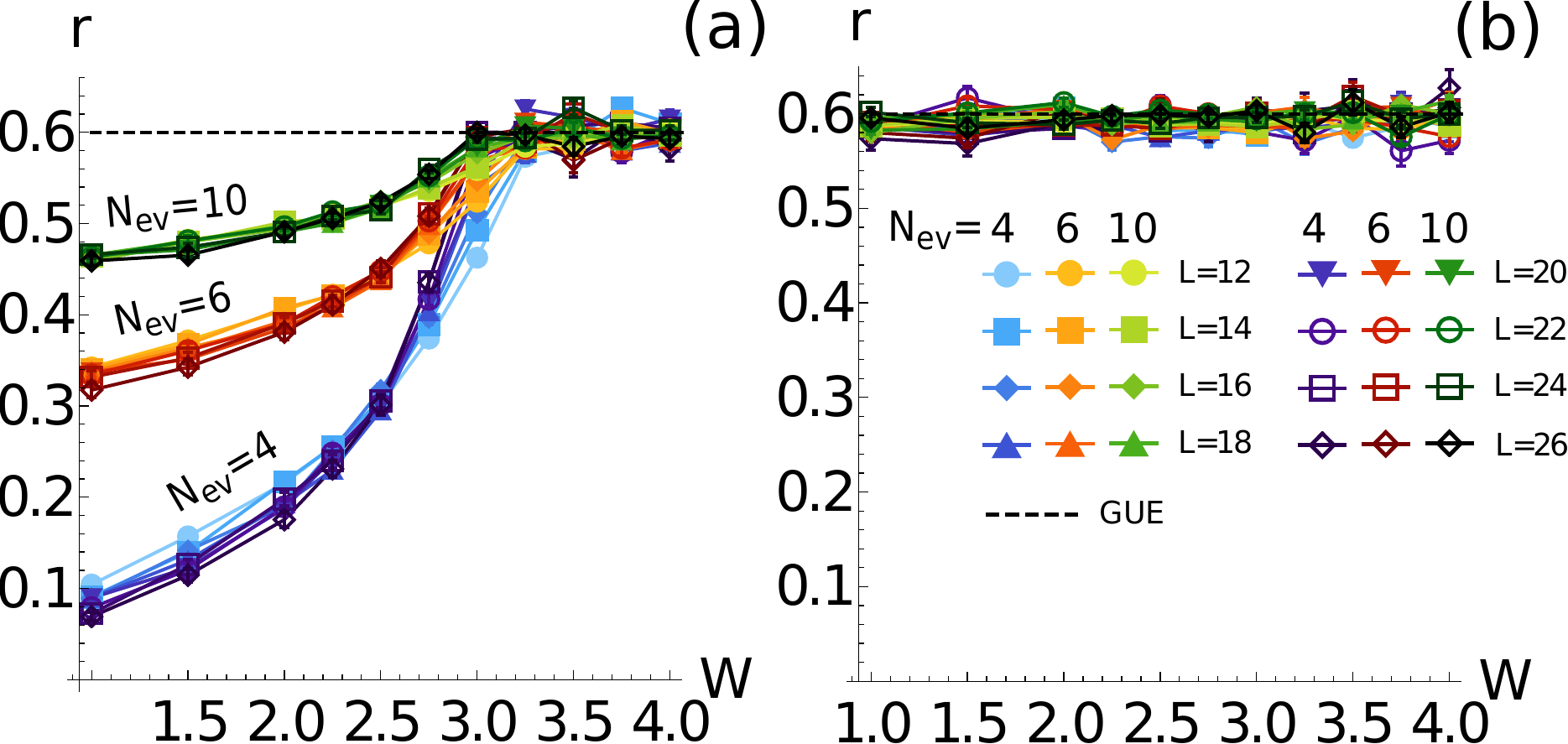}\caption{Level spacing statistics results for the quantity $r$ (defined in
the text) for different linear system sizes $L$ and as a function of the
disorder strength $W$. The results were averaged over $N_{c}\in[250,1000]$
disorder configurations. $N_{{\rm ev}}$ indicates the number of lowest
energy eigenvalues used to compute the spacings. (a) Averaging process
considering the spacing around $E=0$ (between the lowest positive
and highest negative eigenvalues). (b) Averaging process removing
the spacing around $E=0$. The legend in figure (b) also applies to
figure (a).\label{fig:LSS_results}}
\end{figure}

To complement the spectral analysis made with ED, we also studied
the statistics of the energy levels. To do so, we computed the quantity
$r=\langle r_{i}\rangle_{i\in\{N_{E}\},N_{c}}$, where $r_{i}$ is
defined as

\begin{equation}
r_{i}=\frac{\min(\Delta E_{i},\Delta E_{i+1})}{\max(\Delta E_{i},\Delta E_{i+1})}
\end{equation}

\noindent with $\Delta E_{i}=E_{i}-E_{i-1}$ and the average is performed
over the set of lowest energy $N_{{\rm ev}}$ eigenvalues $\{N_{{\rm ev}}\}$
and over $N_{c}$ disorder configurations. The known values \cite{levelstat} for the
quantity $r$ are: (i) $r=0.39$ if the spacings follow a Poisson
distribution; (ii) $r=0.53$ for the Gaussian orthogonal ensemble
(GOE), when the random Hamiltonian does not break time-reversal symmetry;
$r=0.6$ for the Gaussian unitary ensemble (GUE), when the Hamiltonian
breaks time-reversal symmetry. Case (i) applies to ballistic regimes
(due to quasi-integrability) and to insulating regimes (due to energy
level independence). Cases (ii) and (iii) apply to diffusive regimes,
for which Random Matrix theory provides an accurate description.

In the WNL, we expect case (iii) to apply in the MF and SF regimes
due to the usage of twisted boundary conditions that break time-reversal
symmetry. The results are shown in Fig.$\,$\ref{fig:LSS_results}.
In Fig.$\,$\ref{fig:LSS_results}(a) we show the results including
the spacing around $E=0$ (between the lowest positive and highest
negative eigenvalues) in the average to compute $r$. We see that
for $W\geq3$, $r$ follows the GUE value expected for a diffusive
regime, but it takes smaller values for $W\leq2.75$. In this regime, $r$ decreases for smaller $N_{{\rm ev}}$, that is, when we approach $E=0$. We suspect that this
is an effect of the vanishing DOS at $E=0$ in the SM phase, for $W<W_{c}$.
If so, $r$ should follow the GUE value for $W=2.75 > W_c$, which is not observed in Fig.$\,$\ref{fig:LSS_results}(a).
However, for $W=2.75$, $r$ increases with $L$ in constrast with lower $W$, suggesting that it reaches the
GUE value in the thermodynamic limit. In Fig.$\,$\ref{fig:LSS_results}(b),
we remove the energy spacing around $E=0$ and observe that the GUE value is obtained
even in the SM phase. This is expected as the MF regime is diffusive
- the spacings for $E=0^{+}$, where the DOS is finite, should follow
the GUE.

\section{Details on computation of the MF-SF  transition's critical
point}

\label{MFSFcriticalpoint_details}

In this section, we provide additional details on the computation
of the MF-SF critical point, addressed in the main text.

We start by reintroducing the quantity $\phi_{\beta}(W,L)$, defined
as

\begin{equation}
\phi_{\beta}(W,L)=[(\Gamma L)^{-1}+\beta\Gamma ]^{-1}\label{eq:phi_beta_appendix}
\end{equation}

This quantity has the following behavior in the different phases:
\begin{itemize}
\item \textbf{MF: }$\Gamma\sim L^{-1}$, therefore the second term vanishes
with $L^{-1}$ and the first becomes $L-$independent. Furthermore,
the characteristic length scale $\Gamma L P$, with $P$ being the loop's perimeter, is an increasing
function of $W$. As a consequence, $\phi_{\beta}(W,L)$ increases with
$W$ for fixed $L$ and $\beta$;
\item \textbf{SF:} $\Gamma\sim L^{0}$, therefore the first term vanishes
with $L^{-1}$ and the second becomes $L-$independent. 
The characteristic length scale $\Gamma^{-1}$ is a decreasing function of
$W$, and therefore $\phi_{\beta}(W,L)$ decreases with $W$ for fixed
$L$ and $\beta$;
\item \textbf{Critical point: }Both characteristic scales diverge
when $W\rightarrow W_{c}'$, meaning that $\phi_{\beta}(W\rightarrow W_{c}',L\rightarrow\infty)\rightarrow\infty$.
\end{itemize}
We can therefore conclude that the maximum of $\phi_{\beta}(W,L)$,
$W_{{\rm max}}(L)$, for a given system size $L$ and parameter $\beta$
should correspond to the critical point $W_{c}'$ as $L\rightarrow\infty$.

\begin{figure}
\centering{}\includegraphics[width=1\columnwidth]{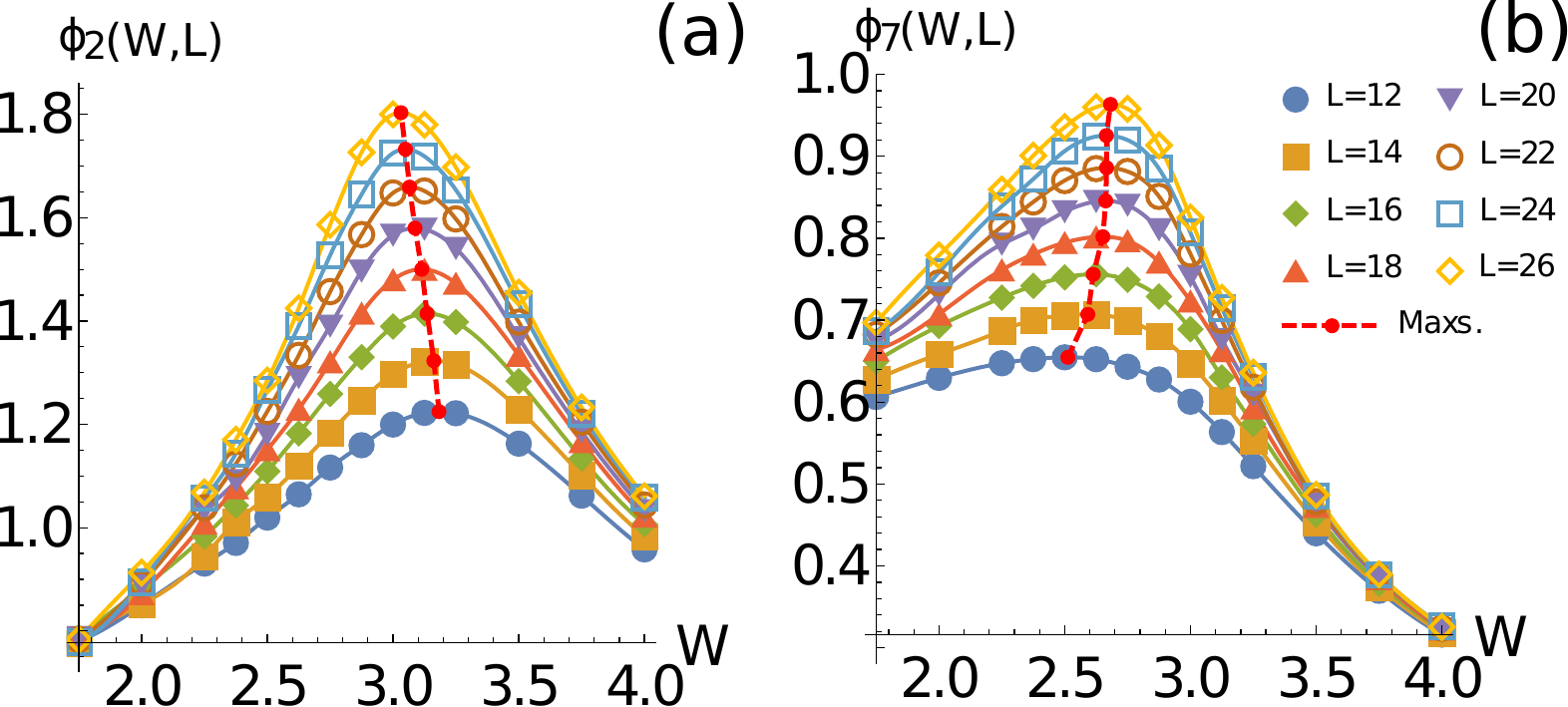}\caption{$\phi_{\beta}(W,L)$ defined in Eq.$\,$\ref{eq:phi_beta_appendix}
for $\beta=2$ (a) and $\beta=7$ (b).\label{fig_Appendix:phi_beta}}
\end{figure}

We now turn to explain the need to choose $\beta$.
In the thermodynamic limit, this factor should have no influence on
the behavior of $\phi_{\beta}(W,L)$. However, that is not true for
finite $L$, as it can be seen in Fig.$\,$\ref{fig_Appendix:phi_beta}.
This makes it more difficult to extract the critical point by studying
the maxima of $\phi_{\beta}(W,L)$. However, we can notice that
there is a curious change in behavior as a function of $\beta$, as
shown in Fig.$\,$\ref{fig4:quantity}(b) of the main text. For smaller
$\beta$, $W_{{\rm max}}(L)$ decreases with $L$, while for larger
$\beta$, it increases. At the critical point, however, $W_{{\rm max}}(L)$
should become constant with $L$ - and therefore the problem of finding
$W_{c}'$ can be reduced to finding $\beta$ such that $W_{{\rm max}}(L)$
becomes $L-$independent.

\begin{figure}
\centering{}\includegraphics[width=1\columnwidth]{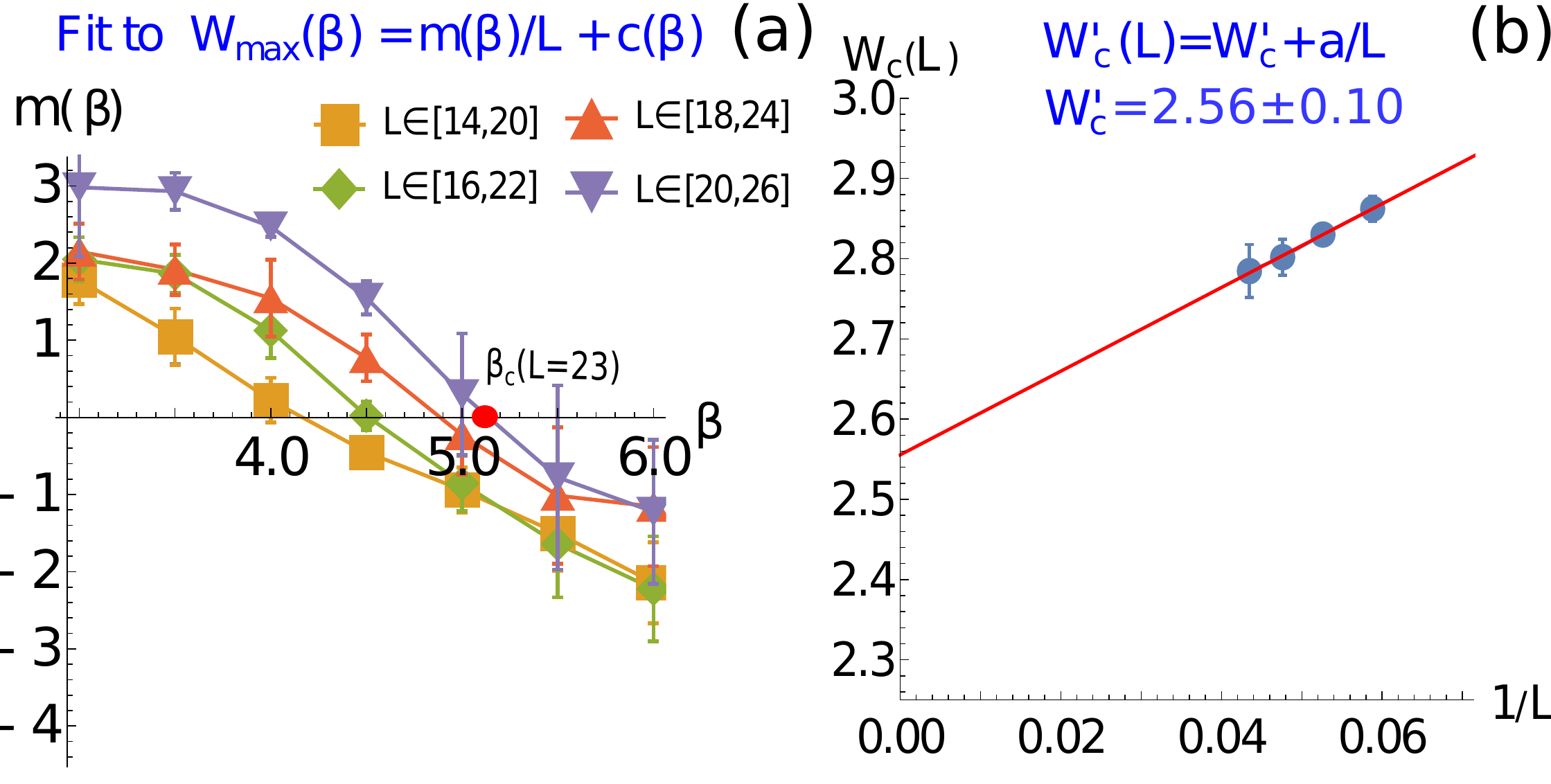}\caption{(a) Parameter $m(\beta)$ used in the fitted model in Eq.$\,$\ref{eq:fit_Wmax_L}, for different ranges of system sizes. The condition $m(\beta_c)=0$ defines $\beta_{c}(L=\langle\{L_{i}\}\rangle)$. (b) Fit of the data $W_c'(L)$ to the model $W_{c}'(L)=W_{c}'+a/L$. \label{fig_Appendix:betac_Wc_computation}}
\end{figure}

There is however an additional difficulty. The value of $\beta$ for
which $W_{{\rm max}}(L)$ becomes $L-$independent depends on $L$
itself. This can be seen clearly in Fig.$\,$\ref{fig_Appendix:betac_Wc_computation}(a).
There, we select groups of 4 consecutive system sizes and fit $W_{{\rm max}}(L)$
to the expression

\begin{equation}
W_{{\rm max}}(L)=m(\beta)/L+c(\beta)\label{eq:fit_Wmax_L}
\end{equation}

When $m(\beta)=0$, $W_{{\rm max}}(L)$ becomes $L$-independent for
a given range of sizes. If we consider for instance $L\in[12,18]$,
the $\beta$ value for constant $W_{{\rm max}}(L)$, $\beta_{c}$,
is $\beta_{c}\approx3.6$, while if we consider $L\in[20,26]$, $\beta_{c}\approx5$.
This of course affects the value of the critical point. The solution
is to find a function $\beta_{c}(L)$ and then a function $W_{c}'(L)$.

The method for finding $\beta_{c}(L)$ is as follows: $\beta_{c}$
is defined through the condition $m(\beta_{c})=0$. The value of $\beta_{c}$
for the set $\{L_{i}\}$ of 4 consecutive sizes is attributed to $L=\langle\{L_{i}\}\rangle$,
that is, the average size of the corresponding set. Then, at $\beta=\beta_{c}(L=\langle\{L_{i}\}\rangle)$,
we can identify $W_{c}'(L=\langle\{L_{i}\}\rangle)\equiv c[\beta_{c}(L)]$.
To extract $W_{c}'(\infty)\equiv W_{c}'$, we can extrapolate $W_{c}'(L)$
to $L\rightarrow+\infty$. To do this, we consider $W_{c}'(L)$ to
be a regular function of $1/L$ and perform a fit to $W_{c}'(L)=W_{c}'+a/L$.
This yields $W_{c}'=2.56\pm0.10$, Fig.$\,$\ref{fig_Appendix:betac_Wc_computation}(b).

To finish this section, we briefly discuss the error analysis in the
$W'_{c}$ computation. To compute the error of $\beta_{c}(L)$, we
obtain the interval for which $|m(\beta_{c}\pm\Delta\beta_{\pm})|-\sigma_{m}\leq0$,
where $\sigma_{m}$ is fitting error of the parameter $m(\beta).$
Then, we compute the error in $\beta_{c}$ through $\sigma_{\beta_{c}}=(\Delta\beta_{+}-\Delta\beta_{-})/2$.

To compute the error in $W_{c}'(L=\langle\{L_{i}\}\rangle)$, we obtain
$c(\beta_{c}\pm\sigma_{\beta_{c}})$ (see Eq.$\,$\ref{eq:fit_Wmax_L})
and define $\sigma_{W_{c}'}=[c(\beta_{c}+\sigma_{\beta_{c}})-c(\beta_{c}-\sigma_{\beta_{c}})]/2$.
Notice that this analysis neglects the fitting errors of $c(\beta_{c})$
and $c(\beta_{c}\pm\sigma_{\beta_{c}})$ because these were computed
to be an order of magnitude smaller than $\sigma_{W_{c}'}$.

\section{Additional details on computation of critical exponents $\nu$ and
$z$}

\label{critical_exps_details}

\subsection{Critical exponent $\nu$}

\begin{figure}
\centering{}\includegraphics[width=1\columnwidth]{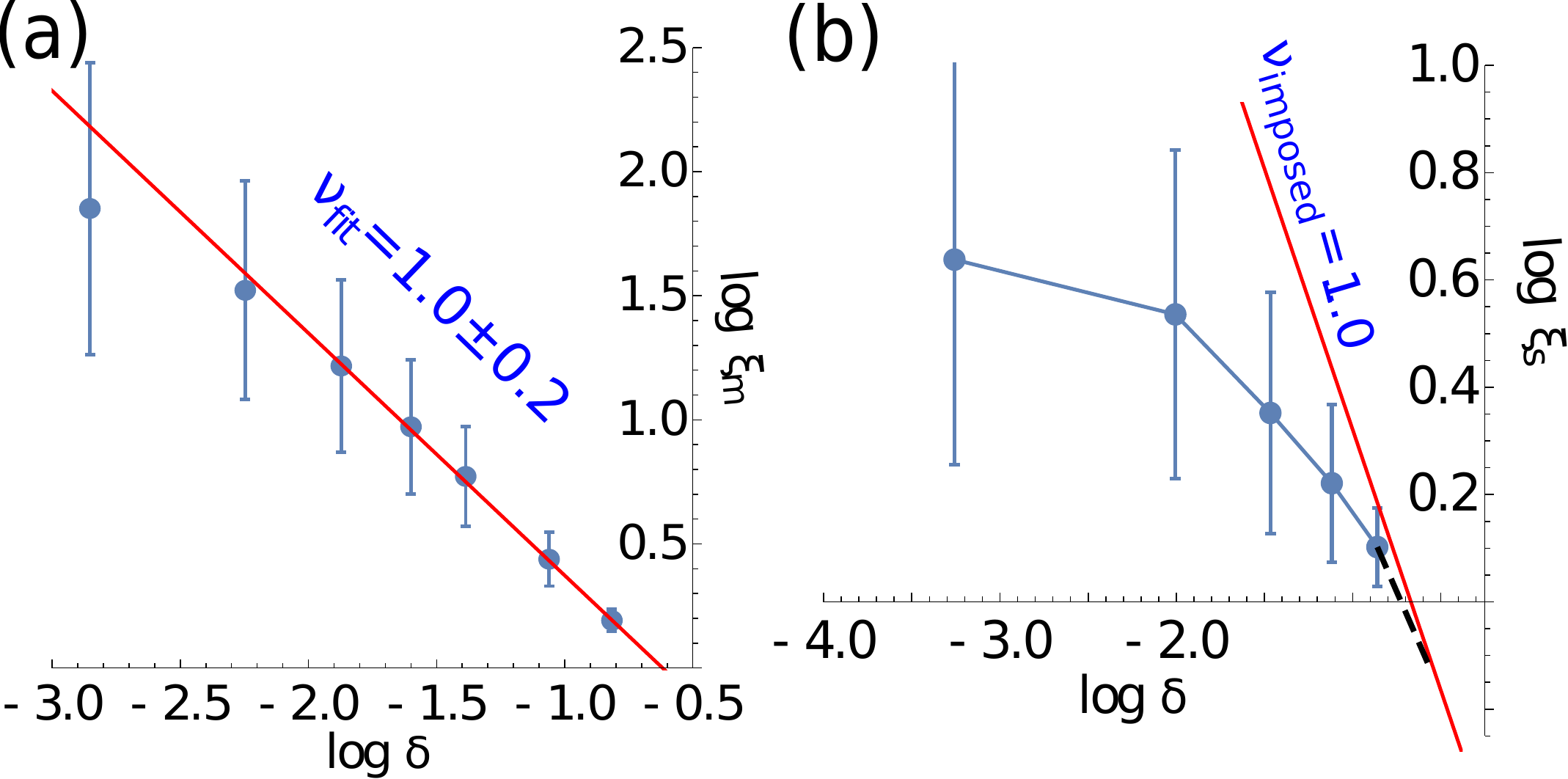}\caption{(a) $\log\xi_{m}$ for $W\in[2.75,3.75]$, considering $W_{c}=2.6$.
The continuous red line corresponds to a linear fit for $W\in[3.125,4]$,
excluding the points near the $W_{c}$, yielding $\nu=1.0\pm0.2$.
(b) $\log\xi_{s}$ for $W\in[1.25,2.5]$, considering $W_{c}=2.6$.
In this case no linear behavior could be identified, even after excluding
the data points in the vicinity of $W_{c}$. As a consequence, it
is not possible to estimate $\nu$ in this case and therefore we show
a line corresponding to the value $\nu=1$, estimated through $\xi_{m}$
in figure (a), along with the data points. The scaling $\xi_{m}\sim\delta^{-1}$
is not incompatible with the data, as the agreement is suggested for the data away from $W=W_c$. \label{fig_Appendix:find_nu}}
\end{figure}

The computation of the critical exponent $\nu$ was carried out through
ED, by obtaining the lowest energy eigenvectors. In particular, it
involved obtaining $\xi_{m}$ and $\xi_{s}$ (up to a constant factor)
by respectively collapsing the curves of the scaling variables $\Gamma$
and $\Gamma^{-1}/L$, as shown in Fig.$\,$\ref{fig:curve_collapsing}
of the main text. By recalling that $\xi_{m}\sim\xi_{s}\sim\delta^{-\nu}$,
with $\delta=|W-W_{c}|/W_{c}$, we can extract $\nu$. We must however
have some caution when using this method. In the vicinity of $W_{c}$,
the correlation lengths are very large and their estimation is associated with a large error. Therefore, we must ignore the close
vicinity of $W_{c}$ to estimate $\nu$. This imposes a problem in
extracting $\nu$ through $\xi_{s}$: we must use data for small $W$,
where resolution issues start to be significant. The determination
of $\nu$ in this case is not, as a consequence, trustworthy. The
results are in Fig.$\,$\ref{fig_Appendix:find_nu}. In (a) we find
$\nu=1.0\pm0.2$ by fitting the $\log\xi_{m}$ versus $\log\delta$ data,
after excluding points in the vicinity of $W_{c}$. In (b), we show
the $\log\xi_{s}$ versus $\log\delta$ data points along with the line with
slope $-\nu=-1$ computed through the fit in (a). In the latter,
we see that the deviations between the slope of the data points and
the slope of the $\nu=1$ line decrease as we move away from $W=W_{c}$.

\subsection{Critical exponent $z$}

To compute the critical exponent $z$, we used the $\rho(E)$ curves
obtained with the KPM. From Eq.$\,$(\ref{eq:scaling_rhoE}), we can
extract $z$ by knowing $\rho(E)$ for $W=W_{c}$.

To ensure that we only used converged data for $\rho(E)$ and for
the used system size ($L=10^{3}$) and number of Chebyshev moments
(up to $N_{m}=2000$), we fitted $\rho(E)$ for $|E|>0.03$ - for
this range of energies, the error between the curves with $N_{m}=1000$
and $N_{m}=2000$ is smaller than $1\%$.

The critical exponent $z$ was computed by fitting $\rho(E)$ at $W=W_{c}$,
with $W_{c}=2.6\pm0.1$.
In order to estimate the error in $z$, $\sigma_{z}$, we must take
into account that it not only depends on the fitting error, but also
on the error in $W_{c}$ and in the energy window used in the fit.
We can compute $z$ for $W=W_{c}\pm\sigma_{W_{c}}$ and then estimate
$\sigma_{z}$ through the difference between $z(W_{c}+\sigma_{W_{c}})$
and $z(W_{c}-\sigma_{W_{c}})$. This is valid if $\sigma_{z}$ computed
in this way is much larger than the fitting error of $z$, which is
the case. Furthermore, we can also vary the energy window used in
the fitting procedure, which also leads to a variation in the value
of $z$. We varied the fitting energy window $E_{{\rm fit}}\in[0.03,E_{{\rm max}}]$
and finally estimated the error as $\sigma_{z}=\max[|z(W_{c}+\sigma_{W_{c}},E_{{\rm max,1}})-z(W_{c}-\sigma_{W_{c}},E_{{\rm max,2}})|/2]$,
with $0.05\leq E_{{\rm max},1},E_{{\rm max},2}\leq0.15$, obtaining
$z=1.9\pm0.1$.

\subsection{Critical exponent $\nu(d-z)$}

To finish this section, we can finally use Eq.$\,$(\ref{eq_Appendix:rho_E_scaling_ThermoLimit})
to see that in the metallic phase and in the thermodynamic limit, we have

\begin{equation}
\rho_{0}\sim\delta^{\nu(d-z)}\label{eq:scaling_dos_metallic}
\end{equation}

Even though $\rho_{0}$ is not converged close to $W=W_{c}$, we can
use the converged data (for larger $W$) to cross-check the results
obtained in the last sections for $\nu$ and $z$ with the scaling
exponent obtained for $\rho_{0}$ in Eq.$\,$(\ref{eq:scaling_dos_metallic}).
By substituting the values already obtained for $\nu$ and $z$, we
get $\nu(d-z)=1.1\pm0.3$.

\begin{figure}[H]
\begin{centering}
\includegraphics[width=1\columnwidth]{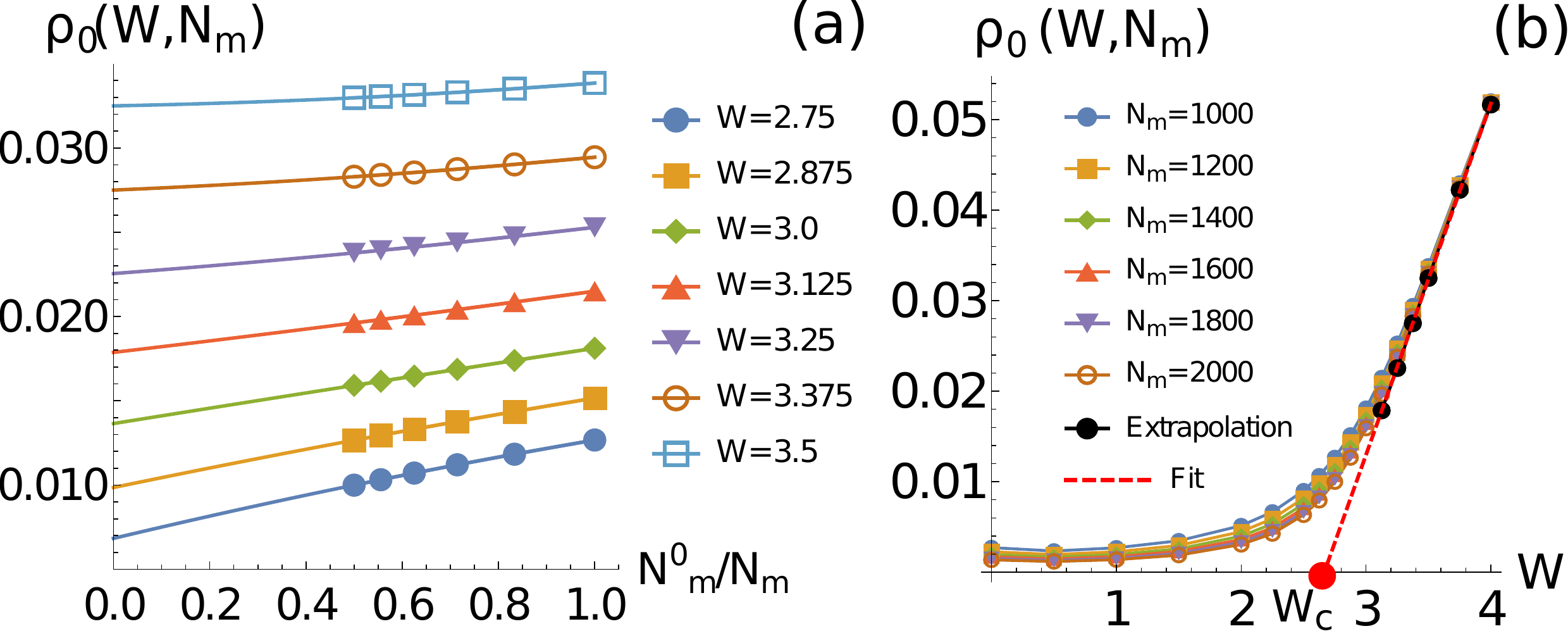}
\par\end{centering}
\caption{(a) Extrapolation of the $\rho(W,N_{m})$ curves for $N_{m}\rightarrow\infty$
by fitting the data to the model $\rho_{0}(W,N_{m})=\rho_{0}(W,\infty)+a/N_{m}+b/N_{m}^{2}$
and identifying $\rho_{0}(W)\equiv\rho_{0}(W,\infty)$. The extrapolation
was considered to be valid as long as the concavity of the fitted
model was positive, that is for $W\protect\geq3.125$. $N_{m}^{0}=1000$
and $N_{m}=2000$ were, respectively, the smallest and largest number
of Chebyshev moments used in the extrapolation. (b) $\rho_{0}(W,N_{m})$
data along with the extrapolation for $N_{m}\rightarrow\infty$, for
$W\protect\geq3.125$. The dashed red line corresponds to the fit
to the model $\rho_{0}(W)=A(W-W_{c})^{\nu(d-z)}$, yielding $W_{c}=2.69\pm0.03$
and $\nu(d-z)=0.97\pm0.03$.\label{fig_Appendix:nu_d-z}}
\end{figure}

In order to obtain $\rho_{0}$ for $W$ closer to $W_{c}$, we extrapolated
$\rho_{0}(W,N_{m})$ for $N_{m}\rightarrow\infty$ by considering
it to be a regular function of $1/N_{m}$. In particular, we fitted
$\rho_{0}(W,N_{m})=\rho_{0}(W,\infty)+a/N_{m}+b/N_{m}^{2}$ and extracted
$\rho_{0}(W)\equiv\rho_{0}(W,\infty)$, considering the extrapolation
to be valid only when the concavity of the fitted model was positive,
that is, for $W\geq3.125$ {[}see Fig.$\,$\ref{fig_Appendix:nu_d-z}(a){]}.
By fitting the model $\rho_{0}(W)=A(W-W_{c})^{\nu(d-z)}$ to the extrapolated
data, we obtained $W_{c}=2.69\pm0.03$ and $\nu(d-z)=0.97\pm0.03$,
compatible with the values obtained before for $W_{c}$, $\nu$ and
$z$ {[}see Fig.$\,$\ref{fig_Appendix:nu_d-z}(b){]}.

\section{Metal-insulator transition}
\label{MItransition}

For larger disorder strengths, the system undergoes a transition between
metallic and insulating phases. To characterize this transition, we
used the transfer matrix method (TMM) \citep{MacKinnon1981,MacKinnon1983,Hoffmann2002}.
The method considers a finite system with a fixed large longitudinal
dimension and a transverse dimension of size $M$ that is varied in
order to compute the localization length $\lambda_{M}$. We computed
the normalized localization length $\Lambda_{M}=\lambda_{M}/M$ as
a function of $M$: if $\Lambda_{M}$ decreases with $M$, the eigenfunctions
are localized in the thermodynamic limit and therefore the system
is an insulator; on the contrary, if $\Lambda_{M}$ increases with
$M$, the eigenstates are extended and the system is a diffusive metal;
a constant $\Lambda_{M}$ signals a critical point separating the
two regimes.

\begin{figure}[H]
\begin{centering}
\includegraphics[width=1\columnwidth]{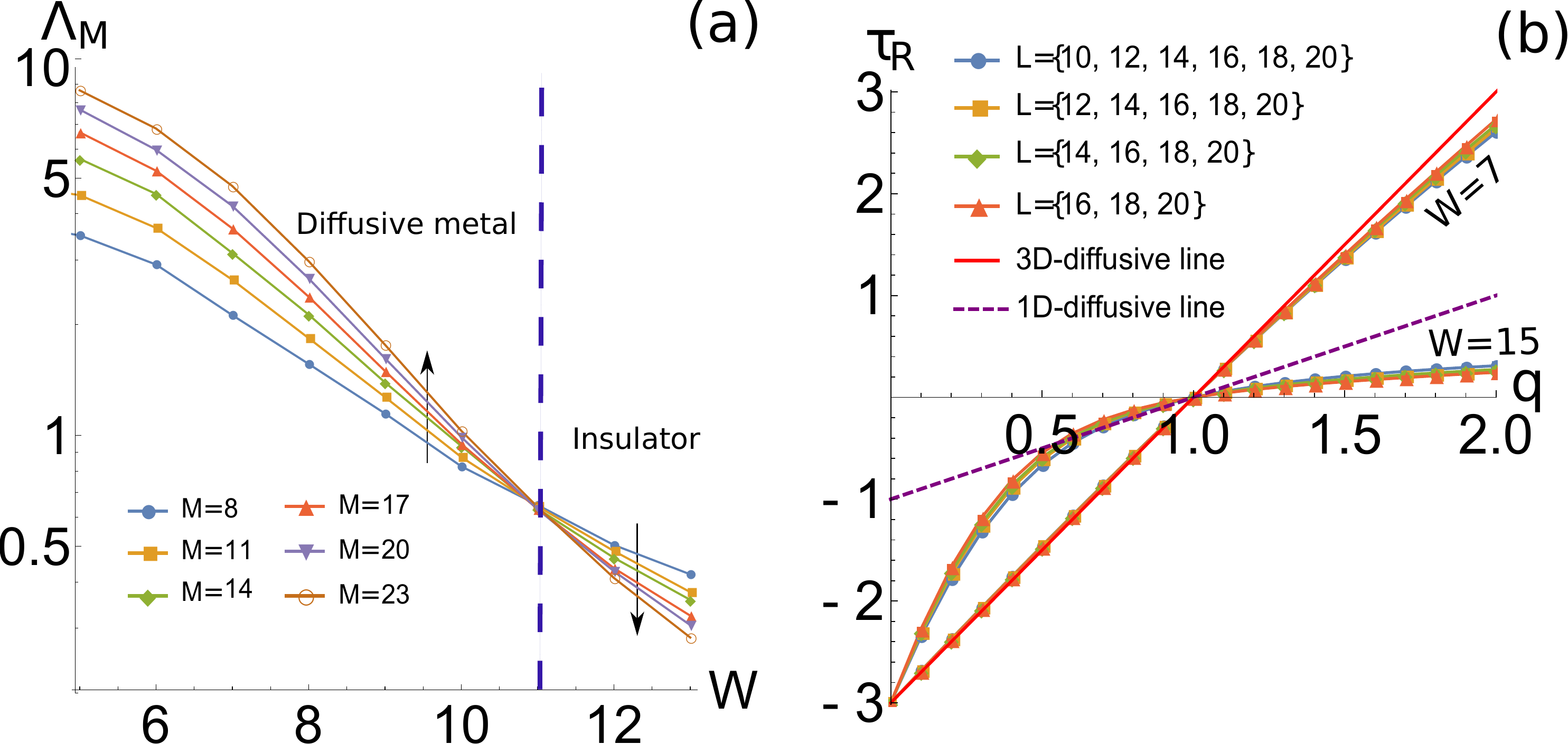}
\par\end{centering}
\caption{(a) Results of the TMM for different transverse sizes $M$. The critical
disorder strength $W_{c}^{l}$ between the metallic and insulating
phases corresponds to the crossing point between the curves of different
system sizes. (b) $\tau_{R}(q)$ exponent for $W=7$ (metallic regime)
and $W=15$ (insulating regime). \label{fig:dif_loc_panel}}
\end{figure}

The precise phase transition point was obtained as the crossing point
between the $\Lambda_{M}(W)$ interpolated curves obtained for different
transverse sizes $M$. The critical point was computed to be $W_{c}^{l}=11.0\pm0.2$.
The results are shown in Fig.$\,$\ref{fig:dif_loc_panel}(a). As
the crossing point between curves of consecutive system sizes oscillated,
we computed $W_{c}$ to be the average of all the computed crossings
and the error to be the corresponding standard deviation.

We can additionally compute the real-space generalized IPR through
\citep{Janssen}

\begin{equation}
\mathcal{I}(q)=\frac{\sum_{\bm{r},\alpha}|\Psi_{\bm{r},\alpha}|^{2q}}{(\sum_{\bm{r},\alpha}|\Psi_{\bm{r},\alpha}|^{2})^{q}}\propto L^{-\tau_{R}(q)}\label{eq:IPRq}
\end{equation}

\noindent where $\Psi_{\bm{r},\alpha}$ is the amplitude of the eigenfunction
at position $\bm{r}$ and sublattice $\alpha$ and $\tau_{R}(q)$
is an exponent. For a diffusive metal, we expect $\tau_{R}(q)=D_{R}(q)(q-1)$,
where $D_{R}(q)$ is the system's dimension. On the other hand, in an
insulator the real-space IPR scales to a constant that provides a
measure of the real-space localization length and therefore $\tau_{R}(q)=0$
for $q>0$.

In Fig.$\,$\ref{fig:dif_loc_panel}(b) we show examples of the $\tau_{R}(q)$
exponent within the metallic ($W=7$) and localized ($W=15$) phases.
In the latter, we do not observe the behavior $\tau_{R}(q)=0$ for
$q>0$. However, when we compute $\tau_{R}$ by fitting the data only
for larger systems, its value increases (decreases) for $q<1$ ($q>1$).
This suggests that in the thermodynamic limit, we should observe the
expected behavior. For $W=7$ we also have some slight deviations
from the 3D-diffusive line for larger $q$ that decrease with system
size. We also computed $\tau_{k}$ (not shown), obtaining the expected
behavior, that is, $\tau_{k}(q)=D_{k}(q)(q-1)$ with $D_{k}(q)=3$.

\section{Wave function for fixed configurations}

\label{wave_fixed_conf}

\begin{figure}[H]
\centering{}\includegraphics[width=1\columnwidth]{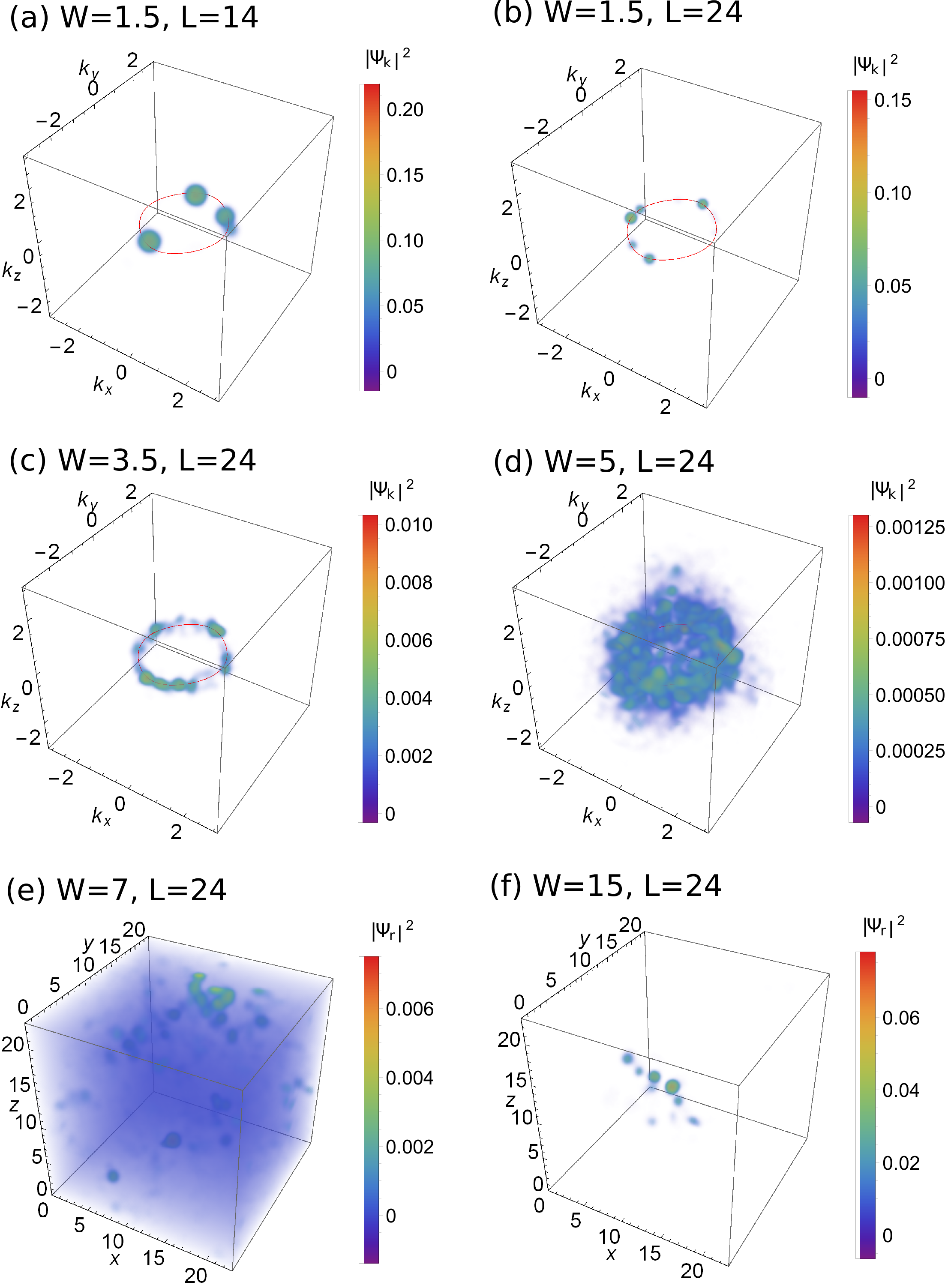}\caption{Plots of the wave function's probability in momentum-space, $|\Psi_{\bm{k}}|^{2}$
(a-d) and real-space, $|\Psi_{\bm{r}}|^{2}$ (e-f) for random configurations.
The color legend corresponds to the probability and on the left of
this legend, we have a transparency legend that varies from black
(completely transparent) to white (completely opaque) - in this way,
only the larger probabilities are observed in the plots. \label{fig:wavefunction_appendix}}
\end{figure}

\begin{figure}[H]
\centering{}\includegraphics[width=0.55\columnwidth]{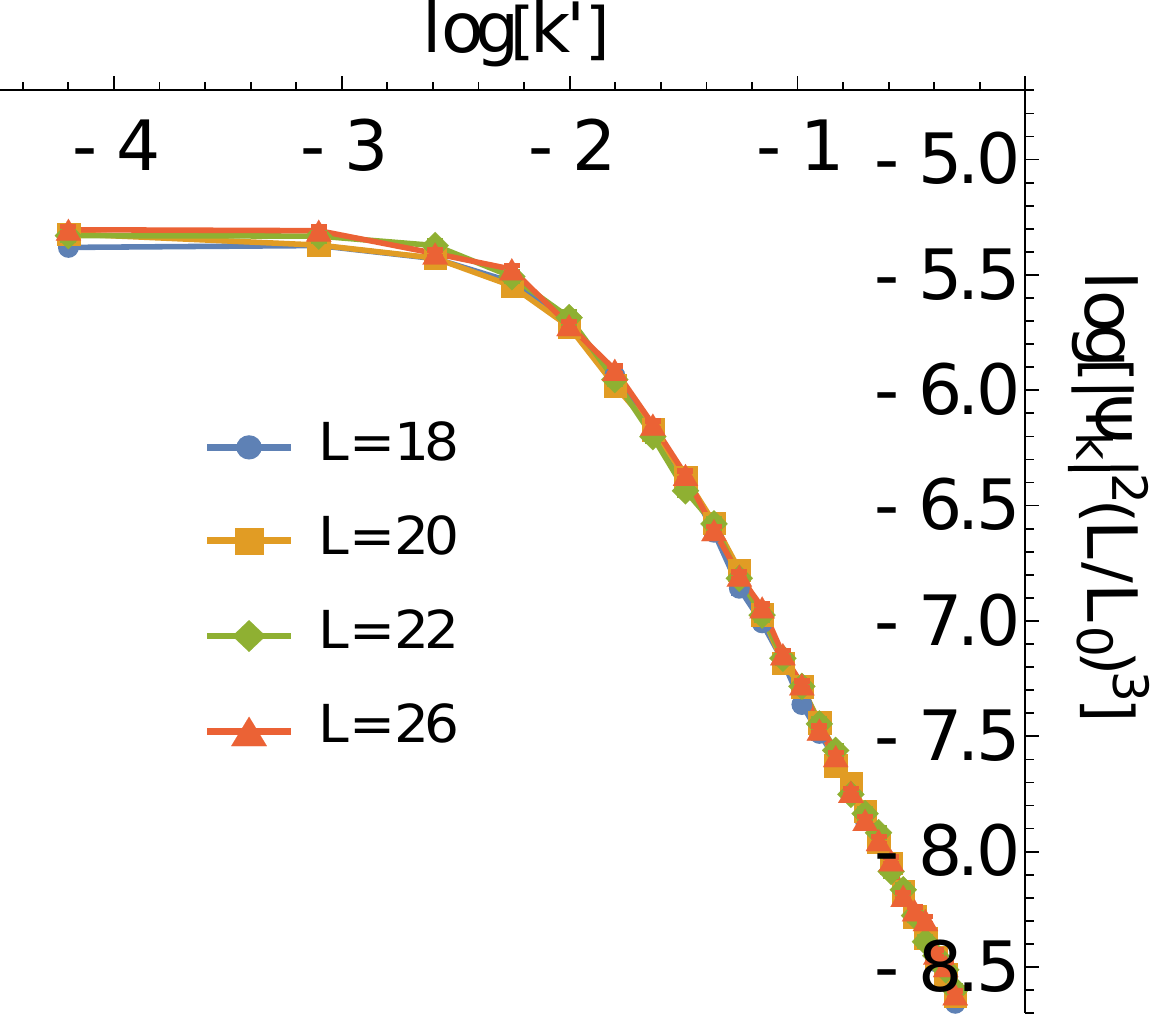}
\caption{$|\Psi_{\bm{k}}|^{2}$ as a function of $k'$ for $W=3.5$ and $k_z=0$, where $k'$ is measured relative to the loop. The system is in the SF regime and the curves of $|\Psi_{\bm{k}}|^{2}$ collapse by rescaling $|\Psi_{\bm{k}}|^{2}\rightarrow|\Psi_{\bm{k}}|^{2}L^{3}$. $L_0=18$ is the smallest used system size.
 \label{fig:wavefunction_appendix_collapse}}
\end{figure}

Even though to carry out an accurate quantitative study it is necessary
to perform an average over a large number of disorder configurations,
it is elucidative to get a picture of the wave function's PD for a
given configuration. In this section we provide some pictures of low-energy
eigenstates for a random configuration and different disorder strengths.
Figs.$\,$\ref{fig:wavefunction_appendix}(a-b) show a case within
the MF regime for $L=14$ and $L=24$. For smaller systems, the wave function
diffuses mostly in the few $\bm{k}$-points that are closer to the
loop. In this regime, the width of the ground-state wave function is
$\Gamma\sim L^{-1}$ and therefore, the average number of points that
have the largest wave function probability increases linearly with
$L$. This can be seen qualitatively by comparing Figs.$\,$\ref{fig:wavefunction_appendix}(a-b)
- for $L=24$, the wave function spreads over a larger number of $\bm{k}$-points.
For $W=3.5$, we have just entered the SF regime and in Fig.$\,$\ref{fig:wavefunction_appendix}(c)
we can see that the wave function spreads around the loop. In this regime, as mentioned in the main text, the wave function collapses by rescaling $|\Psi_{\bm{k}}|^{2}\rightarrow|\Psi_{\bm{k}}|^{2}L^{3}$ (see Fig.$\,$\ref{fig:wavefunction_appendix_collapse}). For $W=5$
{[}Fig.$\,$\ref{fig:wavefunction_appendix}(d){]} we can already
see a large cloud around the loop over which the wave function has
a significant probability. It should be noticed that only the larger
probabilities are being plotted - see description on the transparency
legend in Fig.$\,$\ref{fig:wavefunction_appendix}'s caption. However,
one should not forget that the wave function diffuses over all momentum
space, but the probabilities away from the loop decay as $|\Psi_{\bm{k}}|^{2}\sim k'^{-2}L^{-3}$,
with $k'$ measured relative to the loop.

In Figs.$\,$\ref{fig:wavefunction_appendix}(e-f), we show plots
of the real-space wave function's PD in the metallic ($W=7$) and insulating
($W=15$) phases. In the former, we see that it spreads all over real
space, while in the latter it starts localizing at specific points
in space.

\section{Resolution issues}

\label{sec:Resolution-issues}

To study the wave function's PD in momentum space, we focused on the
plane $k_{z}=0$. Once the nodal loop is located in this plane, there
is an important difference between the resolution that can be attained with respect to the $k_{z}$ direction. In the latter, the resolution is limited by the grid of $k_{z}$ planes, separated by $2\pi/L$. 
However, in the plane $k_{z}=0$, the grid of momentum-space points with different 
$k':\min_{k_{\text{loop}}}\left\Vert \boldsymbol{k}-\boldsymbol{k}_{\text{loop}}\right\Vert  $ grows with $L^2$, providing a better resolution.

\begin{figure}[H]
\centering{}\includegraphics[width=1\columnwidth]{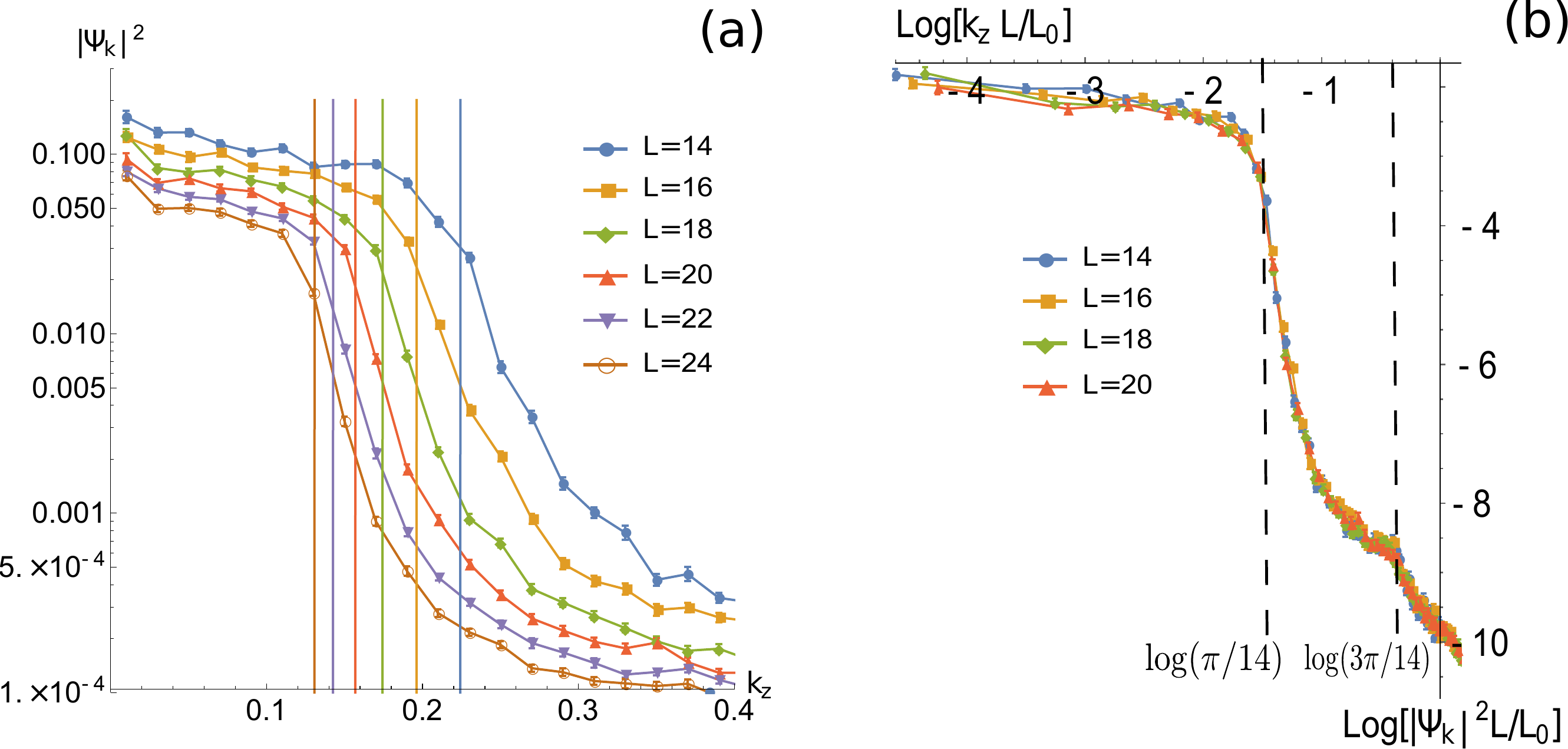}\caption{Wave function's PD $|\Psi_{\bm{k}}|^{2}$ as a function of $k_{z}$
for points $\bm{k}=(k_{x},k_{y},k_{z})$ satisfying $\sqrt{(k_{x}-k_{x}^{loop})+(k_{y}-k_{y}^{loop})}<0.02$,
where $\bm{k}'=(k_{x}^{loop},k_{y}^{loop},0)$ is measured relative to the loop. The results are shown for $W=1.5$. (a) The vertical
lines correspond to $\pi/L$ for the plotted system sizes. (b) The
curves of different sizes are collapsed as $|\Psi_{\bm{k}}|^{2}\rightarrow|\Psi_{\bm{k}}|^{2}L/L_{0}$
and $k_{z}\rightarrow k_{z}L/L_{0}$, where $L_{0}=14$. This figure
shows that resolution problems persist for larger $k_{z}$, as it
can be seen in the peak situated near $k_{z}=3\pi/L$. \label{fig:res.issues.kz}}
\end{figure}

We start by addressing the resolution problems in the $k_{z}$ direction.
As an illustrative example, we show the results for $W=1.5$ in Fig$\,$\ref{fig:res.issues.kz}.
To study this direction independently of the $k_{x}$ and $k_{y}$
directions, we considered only points $\bm{k}=(k_{x},k_{y},k_{z})$
with $\sqrt{(k_{x}-k_{x}^{loop})+(k_{y}-k_{y}^{loop})}<0.02$. We know that the width of the wave function probability in the $k_{z}=0$ plane for this disorder strength is $2\Gamma\approx0.54/L$.
Since there are no significant anisotropies in our model, we expect
the width in the $k_{z}$ direction to be similar, meaning that the
separation between $k_z$ planes is $2\pi/0.54=11.6$ times larger than the width of the
distribution we want to probe (for any system size!). The consequence
is that the distribution assumes a near-plateau for $k_{z}\leq2\pi/L$
and has a sudden drop after this plateau {[}Fig$\,$\ref{fig:res.issues.kz}(a){]}.
These problems persist to larger $k_{z}$, although they become less
significant {[}Fig$\,$\ref{fig:res.issues.kz}(b){]}.

\begin{figure}[H]
\centering{}\includegraphics[width=1\columnwidth]{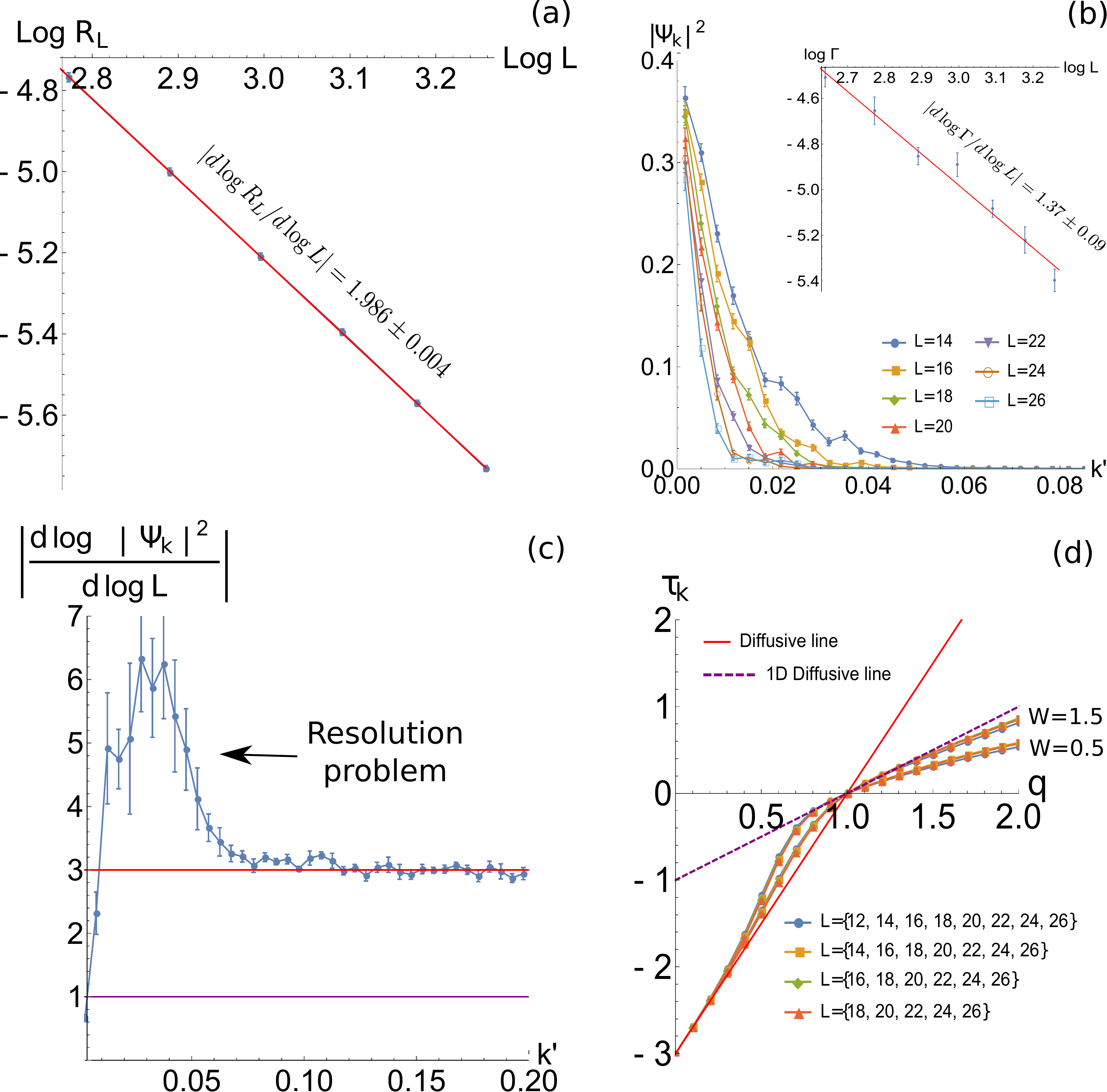}\caption{(a) Log-log plot of the quantity $R_{L}$, defined as the average $\Delta k'$ between consecutive values of $k'$ for a given system size, with $k':\min_{k_{\text{loop}}}\left\Vert \boldsymbol{k}-\boldsymbol{k}_{\text{loop}}\right\Vert  $. A simple fit shows that $|d\log R_{L}/d\log L|=2$, as
expected as a consequence of $R_{L}\sim L^{-2}$ in the plane $k_{z}=0$.
(b) Wave function's PD $|\Psi_{\bm{k}}|^{2}$ as a function of $k'$
for different system sizes, for $W=0.5$. The inset shows that $\Gamma(W,L)\sim L^{-1.4}$.
(c) $|d\log|\Psi_{\bm{k}}|^{2}/d\log L|$ as a function of $k'$.
The fits were made using all the system sizes used in (b). (d) $\tau_{k}(q)$
for $W=0.5$.\label{fig:res_problems_kz0}}
\end{figure}

We now turn to the small disorder resolution problems in the $k_{z}=0$
plane. In this case, although we have better resolution, it is still finite. To see this, we can define the quantity $R_L$ corresponding as the average $\Delta k'$ between consecutive values of $k'$ for a given system size.  We can obtain it simply
by: ordering the $k'$ values obtained for a given
twist in ascending order; computing the spacings $s_{k',i}=k'_{i+1}-k_{i}'$;
averaging the spacings for a given twist, and finally averaging over
twists. The results are shown in Fig.$\,$\ref{fig:res_problems_kz0}(a)
for sizes $L\in[16,26]$ and show that although our ability to resolve the wave function increases with $L^2$, it is still bounded by the value of $R_L$. As shown in the main text, in the MF phase,
we have $\Gamma(W,L)=f(W)/L$ and $f(W)$ increases with the disorder
strength. For small disorder, the attainable system sizes are not
enough to provide a small enough resolution to probe the wave function.
The effects of the lack of resolution start appearing for $W<1.5$
(although still small). We show an example for $W=0.5$, where they
are clear. In Fig.$\,$\ref{fig:res_problems_kz0}(b), we see that
$\Gamma(W,L)\sim L^{-1.4}$ and in Fig.$\,$\ref{fig:res_problems_kz0}(c)
that $|\Psi_{\bm{k}}|^{2}\sim L^{-\mu}$, with $\mu<1$ for $k'\rightarrow0$.
This could suggest erroneously that the wave function spreads over
a fractal dimension $d<1$ for low disorder strengths. However, in
Fig.$\,$\ref{fig:res_problems_kz0}(c) we also observe a sudden drop
in $|\Psi_{\bm{k}}|^{2}$, that can be inspected by $|d\log|\Psi_{\bm{k}}|^{2}/d\log L|$
raising above $3$ for small $k'>0$. If the resolution was small
enough, the curve in Fig.$\,$\ref{fig:res_problems_kz0}(c) should
increase continuously from $|d\log|\Psi_{\bm{k}}|^{2}/d\log L|=1$
($k'<\Gamma$) to $|d\log|\Psi_{\bm{k}}|^{2}/d\log L|=3$ ($k'>\Gamma$).
Indeed, this problematic behavior can easily be reproduced
by sampling a Lorentzian distribution of width $\Gamma\sim L^{-1}$
with $R_{L}>\Gamma$. The lack of resolution results
in an erroneous scaling $\Gamma(W,L)\sim L^{-\eta}$, with $\eta>1$
- in the limit that the $R_{L}\gg\Gamma$, we have $\eta\approx2$.
This problem also has direct consequences in the multifractal analysis:
$\tau_{k}(q)$ becomes smaller (larger) for $q>1$ ($q<1$), as shown in  Fig.$\,$\ref{fig:res_problems_kz0}(d),
which can again suggest, erroneously, the existence of a MF regime
with a fractal dimension $d<1$.

\end{document}